# Applications of Machine Learning in Polymer Materials: Property Prediction, Material Design, and Systematic Processes

Hongtao Guo[1] Shuai Li[2] Shu Li[3,*]

1, School of Electrical and Electronic Engineering, Harbin University of Science and Technology, Harbin, Heilongjiang 150080, People's Republic of China

2, School of Materials Science and Chemical Engineering, Harbin University of Science and Technology, Harbin, Heilongjiang 150080, People's Republic of China

3, Key Laboratory of Engineering Dielectric and Applications (Ministry of Education), School of Electrical and Electronic Engineering, Harbin University of Science and Technology, Harbin, Heilongjiang 150080, People's Republic of China

**Abstract：** This paper systematically reviews the research progress and application prospects of machine learning technologies in the field of polymer materials. Currently, machine learning methods are developing rapidly in polymer material research; although they have significantly accelerated material prediction and design, their complexity has also caused difficulties in understanding and application for researchers in traditional fields. In response to the above issues, this paper first analyzes the inherent challenges in the research and development of polymer materials, including structural complexity and 'the limitations of traditional trial - and - error methods. To address these problems, it focuses on introducing key basic technologies such as molecular descriptors and feature representation, data standardization and cleaning, and records a number of high - quality polymer databases. Subsequently, it elaborates on the key role of machine learning in polymer property prediction and material design, covering the specific applications of algorithms such as traditional machine learning, deep learning, and transfer learning; further, it deeply expounds on data - driven design strategies, such as reverse design, high - throughput virtual screening, and multi - objective optimization. The paper also systematically introduces the complete process of constructing high - reliability machine learning models and summarizes effective experimental verification, model evaluation, and optimization methods. Finally, it summarizes the current technical challenges in research, such as data quality and model generalization ability, and looks forward to future development trends including multi - scale modeling, physics - informed machine learning, standardized data sharing, and interpretable machine learning.

* Corresponding author.

E-mail addresses: lishu@hrbust.edu.cn

## 1 Introduction

As an important branch of material research, polymer science is gradually shifting its research paradigm from traditional experiment - driven to data - driven. The vigorous development of machine learning technology provides strong support for this transformation. In recent years, this technology has made remarkable progress in the fields of polymer material discovery, property prediction, and process optimization, showing broad application prospects. However, how to help researchers in traditional fields understand and apply these rapidly evolving technologies has become a key challenge for promoting the successful transformation of the paradigm. To address this challenge, this study focuses on exploring the application progress of machine learning technologies in polymer research, systematically sorts out their development context and research status, and refines efficient and practical methodologies and systematic processes, aiming to provide valuable references for polymer material researchers to enter this field.

The structure of this review is shown in Figure 1.This study systematically sorts out the application system of machine learning in polymer science: Section 2 elaborates on the data characterization and preprocessing methods of polymer materials, including molecular descriptor construction, data standardization processes, and enhancement technologies; Section 3 comprehensively analyzes the application of various machine learning algorithms in property prediction, covering multi - level technologies such as traditional methods, deep learning, and transfer learning; Section 4 focuses on exploring data - driven polymer design strategies, including innovative methods such as reverse design, high - throughput screening, and multi - objective optimization; Section 5 discusses the key links of experimental verification and model optimization; Section 6 demonstrates practical application results through typical cases; finally, Section 7 summarizes the current challenges and looks forward to future development paths. This review clearly presents the complete knowledge system and technical route of machine learning technology in polymer science research.

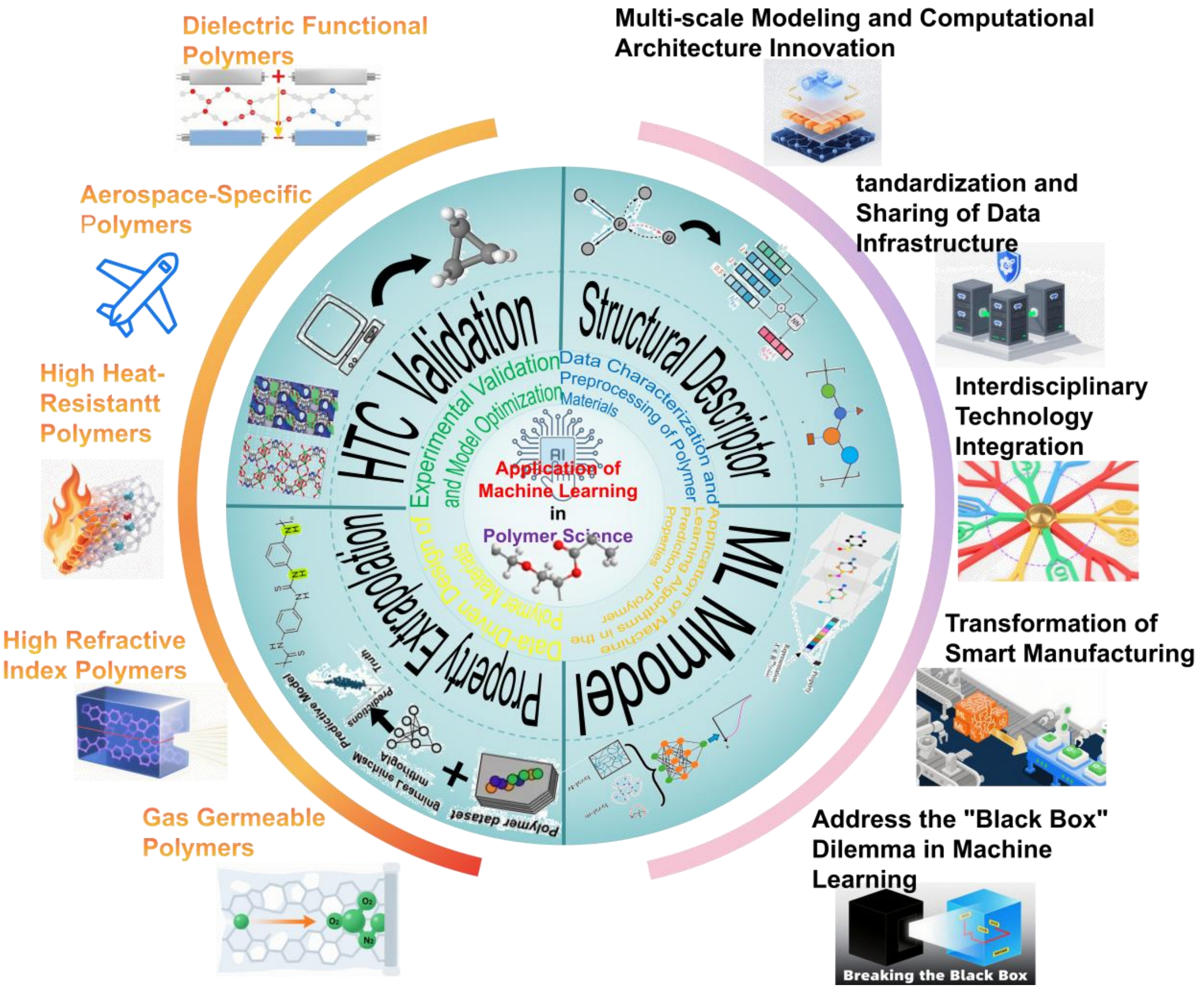

**Figure 1** The figure illustrates a framework composed of four aspects of applying machine learning to polymeric materials: structural descriptors, machine learning models, ML-enabled property extrapolation of polymers, and high-throughput computation. The left and right sides depict existing application cases and the encountered challenges, respectively.

## 1.1 Research Background and Significance

As a basic material in modern industry, polymer materials face long - term challenges in precise design and performance improvement due to their structural complexity and multi - functional requirements[1]. Traditional research methods mainly rely on chemical intuition and trial - and - error methods, which are not only inefficient but also make it difficult to fully grasp the complex structure - property relationships of polymer materials[2]. With the arrival of the big data era, the combination of artificial intelligence and traditional scientific research has given birth to a new paradigm of "AI for Science". As an important branch of artificial intelligence, machine learning has shown significant advantages in revealing the in - depth physical and chemical laws of polymer materials due to its excellent high - dimensional data processing ability[3].

The core challenge in the field of polymer science lies in the fact that the relationship between its huge and complex multi - scale structural characteristics and properties has not been fully mastered. Polymer materials are usually composed of a collection of one or more similar molecules rather than a single structure, which brings unique challenges to traditional chemical representation and machine learning methods[4]. For example, the low thermal conductivity of intrinsic polymers contradicts their wide application requirements in the fields of integrated circuit packaging and organic semiconductors. However, due to the complex synthesis process and high cost of polymers, the publicly available reliable polymer thermal conductivity data are very scarce, which seriously hinders the understanding of the mapping relationship between the micro - structure of polymers and thermal conductivity [5]. Machine learning technology provides a new possibility to solve this problem through its ability to extract useful relationships from limited data [6].

The application of machine learning in polymer science has multiple practical significances. In terms of material design, machine learning can efficiently handle the huge chemical and configuration space of polymers and accelerate the discovery process of new materials[7]. Through the machine learning - assisted inverse analysis method of polymer synthesis, the appropriate polymerization reaction conditions can be quickly and accurately predicted, thereby efficiently developing high - performance polymer materials [8]. In terms of property prediction, machine learning models can handle meaningful patterns in large - scale data that are difficult for humans to interpret, which is particularly useful for systems with complex interactions [9]. Especially when dealing with the complex structure - function relationships of polymer materials, machine learning can establish connections between the chemical composition and conformation of molecular chains, the aggregated structure, and macro - properties [8][10].

From the perspective of industrial application, the introduction of machine learning technology is reshaping the R & D paradigm of polymer materials. The traditional "trial - and - error" experiment has been replaced by the intelligent R & D model of "prediction - verification", which not only changes the working mode of researchers but also redefines the performance boundaries of future energy equipment[11]. In many industries such as aerospace, automobile manufacturing, energy development, and biomedicine, machine learning technology can quickly and accurately predict material properties, significantly shortening the R & D cycle and reducing costs [12]. For example, in the field of polymer composites, machine learning models can solve the thermal management problems that are difficult to handle with traditional development methods by analyzing a large amount of experimental data [13].

The particularity of polymer science also puts forward unique requirements for the application of machine learning. Since polymer materials are usually a collection of one or more similar molecules rather than a single structure, traditional chemical representation methods face challenges[4]. At the same time, the scarcity of high - quality experimental data limits the effectiveness of supervised learning methods, especially in polymer property prediction tasks[14]. These challenges have prompted researchers to develop new methods, such as combining machine learning and high - throughput molecular dynamics simulation to predict material properties[15], and using transfer learning technology to solve the problem of data distribution differences[16].

## 1.2 Research Status

In recent years, the field of polymer science has witnessed the rapid development of machine learning technology, and its application has expanded from basic property prediction to cutting - edge directions such as synthesis optimization and inverse design. In terms of property prediction, the model built by the XGBoost algorithm based on 1774 sets of experimental data can predict 7 key indicators including density and heat distortion temperature at the same time, with an average $R^2$ value as high as 0.95[13]. Deep learning architectures such as hybrid CNN - MLP models and graph convolutional networks have shown excellent performance in predicting properties such as polymer modulus and thermal transition temperature[17].

The field of material design is experiencing a paradigm shift from forward prediction to inverse design. The machine learning platform developed by Chen Mao's team has realized the accurate prediction of polymerization reaction conditions and revealed the mechanism of multi - factor synergy[18]. Deep learning technologies such as GANs and VAEs are used for the design of new compounds, while RFs and GBDTs are widely used for property prediction[19]. The polyBERT model has significantly improved the efficiency of material design by establishing an end - to - end polymer informatics pipeline[20].

In terms of synthesis process optimization, the application of machine learning in free radical polymerization systems has achieved remarkable results, and the experimental data are highly consistent with the prediction results[18]. The development of automated platforms such as RadonPy has promoted the progress of polymer dynamics simulation, and multi - task learning technology has effectively solved the problem of predicting polymer - solvent miscibility. The team of East China University of Science and Technology has realized the accurate prediction of polymer antibacterial activity under small sample conditions, and only 1060 data points are needed to complete the model training[22].

Current research still faces several key challenges. The standardized characterization of biomedical parameters such as degradation time needs to be improved urgently[23], and the complexity of polymer structures makes it difficult for traditional chemical representation methods to accurately describe their sequence and topological characteristics[24]. To address these problems, transfer learning technology and new polymer representation methods are becoming research hotspots[25].

# 2 Data Characterization and Preprocessing of Polymer Materials

Data characterization and preprocessing of polymer materials are key links in the application of machine learning, and their quality directly determines the performance of subsequent models. This process needs to extract valuable information from multi - source data such as experimental measurements, computational simulations, and literature mining, and convert it into structured data suitable for machine learning algorithms through standardized processing. Due to the complex molecular structure, variable physical and chemical properties, and non - linear structure - property relationships of polymer materials, their data characterization faces unique challenges[27]. As shown in Table 1, the key descriptors and their characterization methods of polymer materials in the dimensions of structural features, physical features, chemical features, and multi - scale features provide multi - level characterization tools for understanding the structure - activity relationship and property prediction of polymer materials. Through technical means such as feature engineering and data cleaning, researchers can construct more reliable polymer datasets, laying a solid foundation for subsequent machine learning modeling.

Table 1 Classification and Application Overview of Multi - scale Descriptors for Polymer Materials

| Descriptor | Specific Descriptors | Characterization | Application Scenario |
|---|---|---|---|

| Category | | Method/Source | |
|---|---|---|---|
| Structural Features | Chemical composition of repeating units, bonding mode, sequence arrangement, stereoconfiguration | Coarse - grained representation method [26], BigSMILES [23], curlySMILES [27] | Polymer morphology characterization |
| Structural Features | Degree of polymerization, polydispersity, chain conformation | SMILES combination .modeling [25] | Copolymer system characterization |
| Physical Features | Molecular refractive index, van der Waals surface area | 43 key descriptors extracted by RDKit toolkit [30] | Prediction of physical and chemical properties |
| Physical Features | Atom type, number of bonded hydrogen atoms, atomic degree, implicit valence, aromaticity | Initial atomic feature vector of graph convolutional network [31] | Polymer property learning |
| Chemical Features | Electronic properties, spatial configuration | 434 molecular descriptors extracted by RDKit [30] | Molecular structure analysis |
| Chemical Features | Micro - electronic structure, atomic information, force field parameters | 320 physical descriptors extracted by polymer physical description operators [5] | Polymer system characterization |
| Multi - scale Features | Atomic - level (155), segment - level (197), molecular chain - level (59) descriptors | Three - layer structure characterization method [10] | Dielectric constant research |
| Multi - scale Features | Atomic scale (108), QSPR level (99), morphological description (22) | Ramprasad three - layer characterization method[10] | Polymer material characterization |

## 2.1 Molecular Descriptors and Feature Representation

The numerical characterization of polymer structures is the key basis for the application of machine learning in polymer science, and its core challenge lies in how to convert complex chemical structures into mathematical expressions that can be processed by computers. Polymer chains are usually composed of a large number of small organic molecule units connected repeatedly through covalent bonds, and their micro - structural features include multiple dimensions such as the chemical composition, bonding mode, sequence arrangement, and stereoconfiguration of repeating units [26]. The diverse structures of synthetic polymers (like composition, architecture, and sequence) lead to complex structure - property relationships, posing challenges in soft material design. To tackle this, researchers have developed molecular descriptors and feature representation methods. These methods, such as BigSMILES and ECFP, convert polymer structural features into computable descriptors. By doing so, they enable the mining of structure - property relationships from high - dimensional data, which is crucial for guiding iterative library design and predictive modeling of material properties[32].

SMILES (Simplified Molecular Input Line Entry System) and its extended forms have important application value in the characterization of polymer structures[32]. Although the traditional SMILES syntax has been widely accepted, it is difficult to accurately describe the complex structural features of polymers. For this reason, researchers have successively developed extended representation methods such as BigSMILES [33] and curlySMILES [34]. These methods can more effectively characterize different polymer morphologies such as linear, branched, random, block, alternating, and grafted[23]. Among them, BigSMILES captures the unique chemical properties of polymers by extending the SMILES syntax and shows obvious advantages in dealing with multi - repeating composite units or complex architectures[27]. For copolymer systems, the method of combining SMILES of each repeating unit is usually used for modeling, and structural descriptors such as degree of polymerization, polydispersity, and chain conformation are introduced to improve the characterization[25].

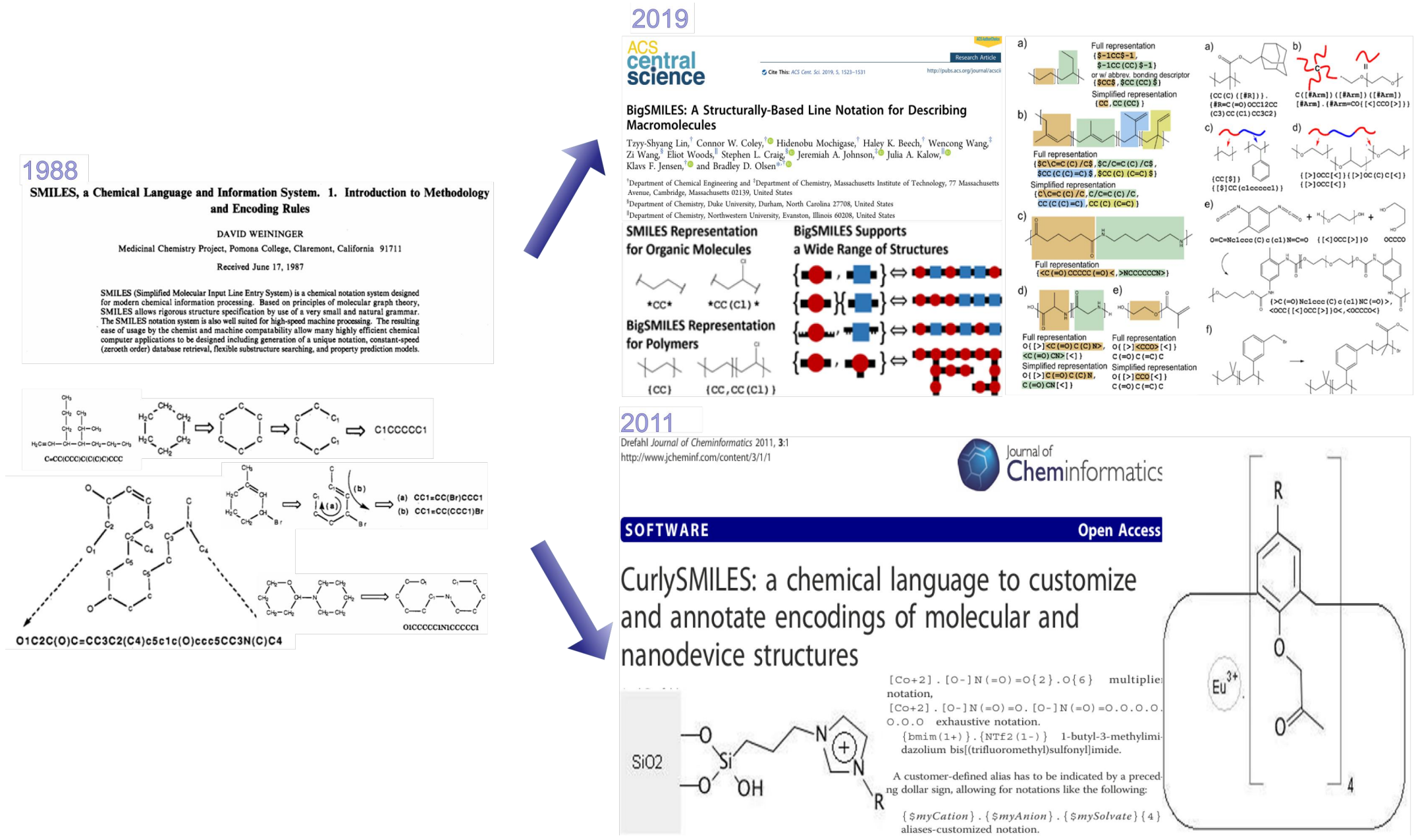

**Figure 2** The figure shows a schematic diagram of the development process of SMILES (Simplified Molecular Input Line Entry System) and its extended forms (BigSMILES, CurlySMILES). From the proposal of traditional SMILES in 1988 [33], to the subsequent development of extended representation methods such as BigSMILES (2019) [35] and CurlySMILES (2011) [36] by researchers to solve its shortcomings in the characterization of complex polymer structures, these methods can more effectively characterize different polymer morphologies such as linear and branched, contributing to the accurate description of polymer structures.

In the field of molecular fingerprint technology, Morgan fingerprints characterize molecular features by identifying all possible substructures, and their improved version MFF further considers the frequency of substructures [28]. Extended Connectivity Fingerprints (ECFP), as one of the commonly used methods, can effectively capture the key substructures and their distribution characteristics in polymers by converting the monomer chemical structure into a binary descriptor vector [29]. In practical applications, researchers use the RDKit chemical information toolkit to conduct in - depth analysis of the molecular structure encoded by SMILES, and can extract 434 molecular descriptors covering dimensions such as electronic properties, spatial configuration, and physical and chemical properties. After screening through Pearson correlation coefficient analysis, 43 key descriptors are finally retained, including core parameters such as molecular refractive index and van der Waals surface area [30].

The graph representation method provides a new research idea for the characterization of polymer structures. Graph Convolutional Networks (GCN) learn polymer properties by iteratively updating node feature vectors, and their initial atomic feature vectors are composed of information such as atom type, number of bonded hydrogen atoms, atomic degree, implicit valence, and aromaticity [31]. Another new method is the graph - based molecular set representation combined with the Weighted Directed Message Passing Neural Network (wD - MPNN) architecture, which captures the average graph structure features of repeating units by parameterizing the description of the underlying molecular distribution [4]. For complex polymer systems, researchers have developed polymer physical description operators and recursive screening optimization processes. 320 physical descriptors are extracted from the micro - electronic structure, atomic information, and force field parameters of the monomer structure. Through the analysis of various statistical parameters and 100 random sequence feature screenings, the dimension is finally reduced to 20 optimized descriptors [5].

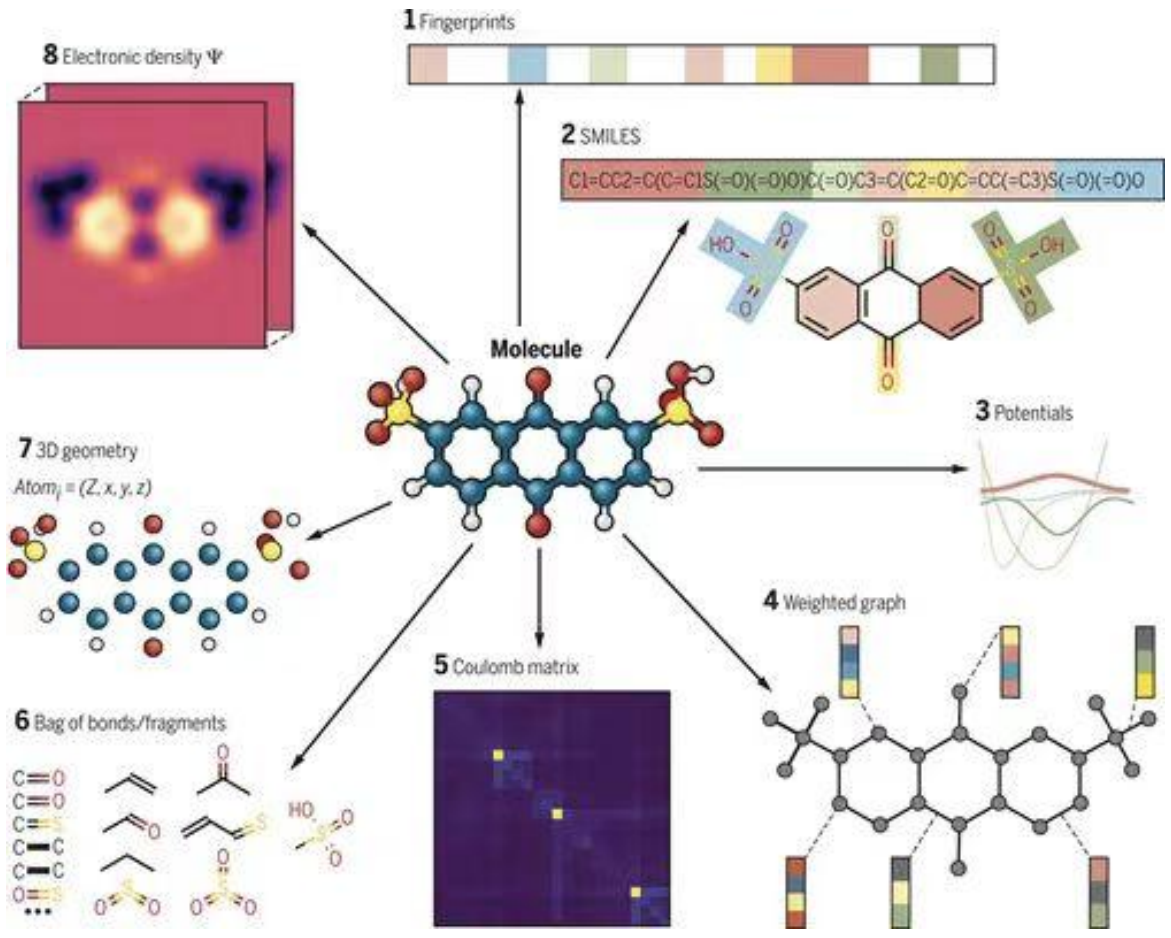

**Figure 3** Different types of molecular representations for the same molecule [54]. (1) Fingerprint vector; (2) SMILES string; (3) Potential energy function; (4) Weighted graph of atoms and bonds; (5) Coulomb matrix; (6) Combination of bonds/fragments; (7) 3D geometry of atomic charges; (8) Electronic density.

The selection of polymer structure descriptors needs to take into account multi - scale features. When studying the dielectric constant, Chen et al. divided the structure into three levels: atomic level, segment level, and molecular chain level, and selected 155, 197, and 59 structure descriptors respectively[10]. Ramprasad et al. adopted a similar three - layer structure characterization method: 108 descriptors are selected at the atomic scale (such as O1 - C3 - C4 segments); 99 descriptors are selected at the Quantitative Structure - Property Relationship (QSPR) level (such as van der Waals surface area); 22 descriptors are selected at the morphological description level (such as the shortest topological distance between rings)[10]. This layered description strategy can comprehensively capture the multi - scale features of polymer materials and provide more abundant structural information for machine learning modeling.

## 2.2 Data Standardization and Cleaning

The standardized processing and quality control of polymer data are the basic links of machine learning modeling, and their quality directly determines the prediction performance of the model. High - quality data is the prerequisite for avoiding the phenomenon of "garbage in, garbage out", which makes data standardization and cleaning a necessary step to ensure the reliability of the model[6]. A major challenge currently facing the field of polymer research is that due to differences in experimental methods and data analysis, datasets from different sources often have compatibility problems and lack uniform standards, which highlights the importance of data preprocessing in the application of machine learning[23]. The characterization data of polymer materials are usually presented in statistical indicators such as molecular weight and its distribution, which further increases the complexity of data processing [26].

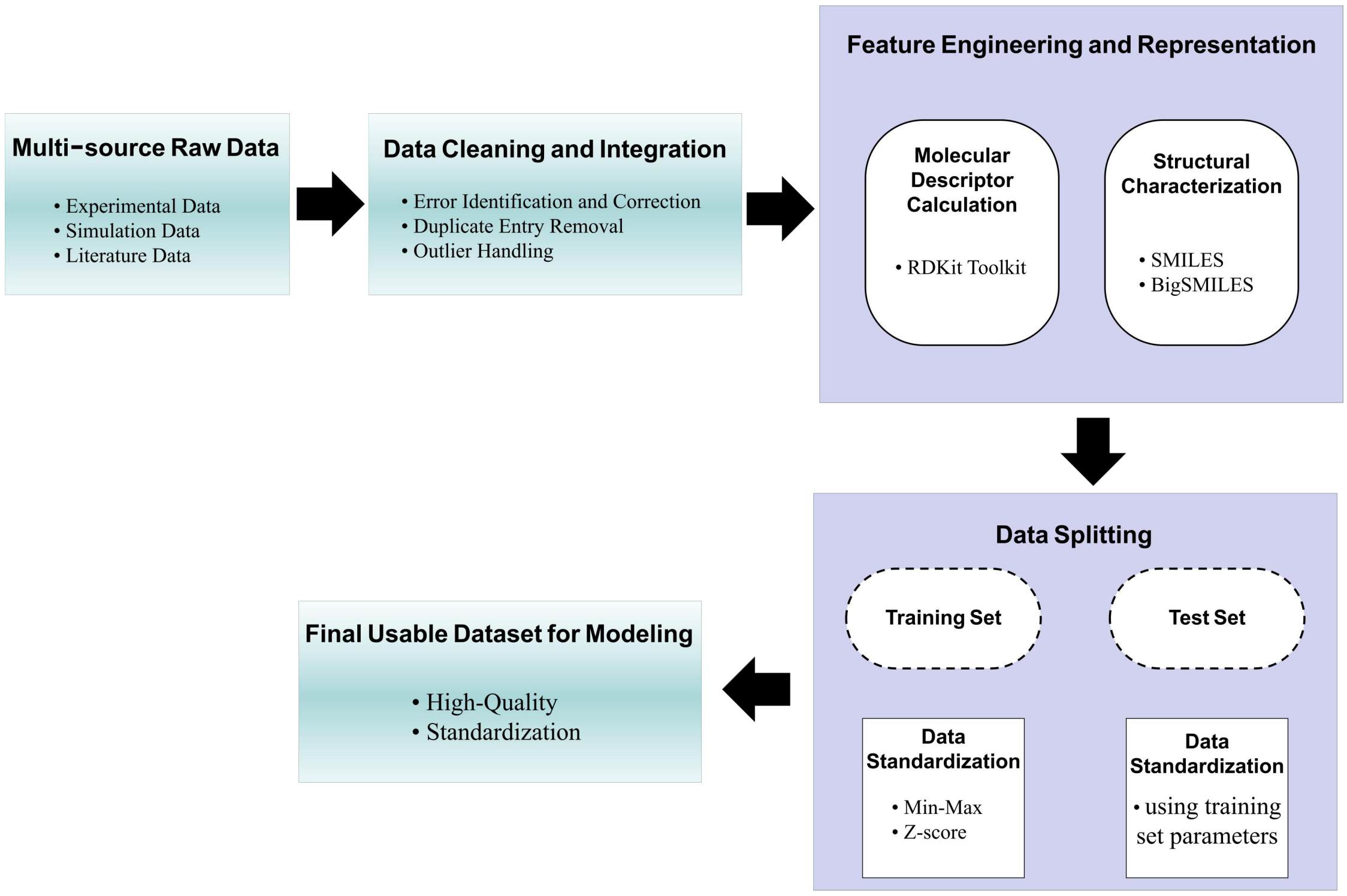

Figure 4 This diagram outlines the systematic preprocessing of polymer data for machine learning.

Data preprocessing mainly includes key links such as error identification and correction, duplicate entry deletion, and outlier handling. In specific operations, it is necessary to standardize or scale the input variables to ensure the consistency of the data range, and at the same time convert categorical variables into machine - readable encoding forms, such as one - hot encoding or label encoding [35]. Feature scaling is an important part of data preprocessing, among which the min - max normalization technology is widely used because it can maintain the uniformity of data distribution [36]. In the actual modeling process, researchers usually divide the training set and test set in a ratio of 8:2 or 9:1, and standardize the two sets of data separately [41].

The dispersion and insufficient standardization of polymer data are the main factors restricting the application of machine learning algorithms. To address this challenge, the polymer research community is developing new database systems, such as PoLyInfo and CRIPT, aiming to realize the effective storage and utilization of polymer data[9]. Among them, the Polydat framework can record structural data and characterization parameters to promote the standardized integration of data; BigSMILES, as an extension of the SMILES format, is specially used to describe the characteristics of repeating units and branch structures of polymers[37]. The PoLyInfo database has now included property data of about 100 polymers, including key properties such as glass transition temperature, melting point, density, and thermal conductivity. These data have undergone strict cleaning and standardization processing, which has significantly improved the prediction accuracy of the model[38]. Table 2 records several commonly used polymer datasets and the property data they record.

Table 2 Commonly Used Polymer Datasets

| **Dataset Name** | **Contained Data** | **Description** | **Web** |
|---|---|---|---|
| **Polymer Genome Platform** | **Refractive Index (RI), dielectric properties, glass transition temperature (Tg)** | **Experimental data repository with 500+ polymer measurements for real-time property prediction** | **https://polymergenome.ecust.edu.cn/** |
| **Khazana** | **Computational materials data** | **Georgia Tech database for machine learning applications in polymer science** | **https://khazana.gatech.edu/dataset/** |
| **Dortmund Database** | **Polymer thermophysical properties** | **Commercial reference database for thermal characteristics** | **https://ddbst.com/** |
| **PoLyInfo** | **Multiscale polymer performance** | **NIMS Japan comprehensive polymer repository** | **https://polymer.nims.go.jp** |
| **NIST Spectral Database** | **Synthetic polymer MALDI mass spectrometry** | **Spectral analysis database for polymer characterizatio n** | **https://maldi. nist.gov** |
| **CROW Polymer Database** | **Physical/mechan ical/ thermal/electric al properties** | **Broad-spectru m polymer properties reference** | **http://polymerdatabase.com** |
| **Material Properties Database** | **Comparative material metrics** | **Industrial materials benchmark including polymers** | **https://www.makeitfrom.com** |
| **Mechanical Properties Dataset** | **Young's modulus, tensile strength, elongation (429 points)** | **Combined literature/MD simulation data for structure-prop erty modeling** | **https://www.kaggle.com/datasets/purushotta mnawale/materials** |
| **Thermal Conductivity Dataset** | **Polymer chain descriptors, DFT calculations** | **Structure-ther mal property relationships for novel polymer design** | **https://researchdata.edu.au/thermal-conduc tivity-dataset/3431817** |
| **Compatibility Dataset** | **Polymer-polyme r interaction** | **Literature-min ed** | **https://github.com/cloudflare/workers-sdk/is sues/193** |

| Dataset Name | Contained Data | Description | Web |
|---|---|---|---|
| | data (1,000+ points) | classification data for blend miscibility | |
| Dielectric Multi-task Dataset | Permeability/diffusivity/ solubility parameters | Fusion of high-fidelity experimental and low-fidelity simulation data | https://github.com/easezyc/Multitask-Recommendation-Library |
| Refractive Index Dataset | Hierarchical fingerprint data for 500 polymers | Multi-scale structural descriptors (atomic/segment/chain level) | https://refractiveindex.info/ |
| PI1M | Polymer structures, synthetic accessibility score | PI1M has ~1M polymers and Schuffenhauer's SA scores , a polymer informatics benchmark. | https://github.com/RUIMINMA1996/PI1M |
| Polymer Genome | Bandgap, dielectric constant, refractive index, atomization energy, Tg, solubility parameter, density | Polymer Genome has computational & experimental polymer data for informatics and property prediction | https://www.polymergenome.org |
| Polymer Property Predictor and Database | Flory-Huggins chi parameters, glass transition temperature (Tg) | A literature-extracted polymer database with chi parameters and Tg, for polymer informatics research | https://pppdb.uchicago.edu |
| Polymer Science Learning Center Spectral Database | Polymer FTIR, Raman, NMR spectra | Experimental spectral database with polymer-specific spectra for identification and structural analysis | https://pslc. uwsp.edu |

The standardized processing of polymer data usually adopts methods such as min - max scaling and z - score standardization to ensure the scale consistency between different features. Data enhancement technologies such as adding Gaussian noise are also used to improve the generalization ability of the model[19]. Since the original polymer data often has non - standardization problems, the cleaning process needs to focus on the identification of data deviations, outlier detection, and standardized processing[39]. In the data screening link, researchers usually exclude polymer structures with a standard deviation exceeding the set threshold, and the thresholds for glass transition temperature and melting point are usually set to 30K[31].

Data management is the primary link in the machine learning - assisted polymer design framework, involving the systematic organization of historical data and new data. When constructing a high - quality initial dataset, converting the polymer structure into a computer - readable form is the basic work [40]. However, problems such as missing reaction parameters and incomplete characterization conditions commonly existing in open - source databases and literature bring significant challenges to the collection of standardized data[41]. The establishment of an initial dataset that conforms to the FAIR principle is crucial to ensuring the reliability of machine learning modeling, which needs to be achieved through systematic experimental data accumulation or high - throughput methods[16].

## 2.3 Data Enhancement Technology

The problem of data scarcity in the field of polymer science seriously restricts the performance improvement of machine learning models. To solve this bottleneck, researchers have developed a variety of innovative data enhancement methods using the group contribution method, as in the research of Ning Liu et al. The physical modeling method simulates the cone calorimeter experiment through the Fire Dynamics Simulator (FDS), generates data on ignition time and peak heat release rate with physical consistency, and effectively expands the training sample library [14]. This method not only avoids the difficulty in obtaining experimental data but also ensures the reliability of the generated data.

In the research of thermal conductivity prediction, transfer learning technology has shown significant advantages. Researchers trained 1000 pre - trained neural network models based on the PoLyInfo and QM9 databases, and then fine - tuned them with limited target data, successfully improving the prediction accuracy[43]. The polyBERT model adopts a molecular fragment recombination strategy, decomposes known polymers into fragments and then recombines them, generating 100 million hypothetical PSMILES strings, which greatly expands the scale of the dataset[44]. This chemical knowledge - based enhancement method not only ensures the amount of data but also maintains the rationality of molecules.

To address the small sample problem, the Bootstrap resampling technology expands 180 experimental samples to 1500 samples, effectively solving the problem of insufficient data in the research of natural fiber - reinforced polymer composites[17]. The graph grammar distillation framework innovatively decomposes the amino acid structure into molecular graph grammar fragments, and realizes the accurate exploration of the high - dimensional polymer space through recombination[22]. These methods all retain the statistical characteristics of the original data well.

The application of generative recurrent neural networks in the PI1M database has generated about 1 million theoretical polymer data[45], and the large language model for polymer property prediction has constructed an extended dataset containing four types of tasks[46]. These data fusion methods significantly increase the amount of data while ensuring quality.

The research team integrated multi-source small molecule databases, generated massive hypothetical structures of 8 polymer types and 1 copolymer type via rule-based polymerization reactions, analyzed the structural characteristics using t-SNE and SA scores, and predicted the thermal, mechanical, and gas permeation properties with a customized FNN model. The study confirmed the performance advantages of hypothetical polymers (especially polyimides), providing support for data-driven polymer research and development[47].

In research where data acquisition is costly, the combination of active learning and Bayesian optimization realizes the efficient utilization of data[25]. At the same time, the collaborative application of high - throughput computing and experiments, through the combination of molecular dynamics simulation and automated experiments, has constructed a high - quality standardized dataset[26]. These multi - source data integration strategies provide systematic solutions for polymer material research.

# 3 Application of Machine Learning Algorithms in Polymer Property Prediction

In recent years, the field of polymer material property prediction has undergone a paradigm shift due to the introduction of machine learning technology. The construction of data - driven models not only accelerates the process of material discovery but also opens up new ways for property prediction. Current research mainly focuses on three technical directions: traditional machine learning methods extract key parameters of molecular structures through feature engineering; deep learning technology uses neural networks to automatically learn the non - linear relationship between material components and properties; transfer learning methods solve the prediction problem under small sample data through knowledge transfer. As shown in Table 2, these algorithm systems have their own advantages in terms of predicted performance indicators and applicable scenarios. The systematic comparison results provide empirical evidence for materials science researchers to select appropriate artificial intelligence methods. These algorithms together form a mapping bridge from molecular features to macro - properties, providing a quantitative theoretical basis for material inverse design.

Table 3 Performance Comparison of Different Machine Learning Algorithms in Material Property Prediction

| Algorithm Category | Representative Model | Predicted Performance Indicator | Applicable Scenario | Literature Reference |
|---|---|---|---|---|
| Traditional Machine Learning | Support Vector Machine (SVM) | Polymer Tg prediction $R^2$=0.91 [10] | Small sample, high - dimensional dataset analysis [6] | [6][10] |
| Traditional | Random Forest (RF) | Thermal conductivity | Processing long input features | [30][50] |

| Algorithm Category | Representative Model | Predicted Performance Indicator | Applicable Scenario | Literature Reference |
|---|---|---|---|---|
| Machine Learning | | prediction $R^2$=0.97 [30] | and noisy data [50] | |
| Traditional Machine Learning | XGBoost | Concrete strength prediction $R^2$=0.98 [49] | Automatically identifying feature interaction relationships [49] | [49] |
| Deep Learning | Graph Neural Network (GNN) | Tg prediction RMSE=30K, $R^2$=0.90 [31] | Processing molecular graph structure data [4] | [4][31] |
| Deep Learning | Transformer | PSMILES processing 100 times faster [44] | Chemical language model construction [44] | [44] |
| Deep Learning | Physics - Informed Neural Network | Thermal conductivity anisotropy prediction [56] | Multi - scale modeling [56] | [56] |
| Transfer Learning | Sim2Real strategy | Thermal conductivity prediction MAE=0.024W/mK [57] | Data - scarce scenarios | [57] |
| Multi - task Learning | polyBERT | Multi - attribute joint prediction [44] | Mining associations between attributes [39] | [39][44] |

### 3.1 Traditional Machine Learning Methods

In the field of polymer science, traditional machine learning algorithms such as Support Vector Machine (SVM) and Random Forest (RF) occupy an important position in property prediction research due to their excellent non - linear modeling capabilities and stability under small sample conditions. Support Vector Machine completes classification and regression tasks by constructing an optimal hyperplane in the high - dimensional feature space, and is particularly suitable for handling the complex mapping relationship between polymer structures and properties. The SVM model using the Gaussian radial basis function as the kernel function has achieved remarkable results in the prediction of polymer glass transition temperature (Tg) and electrostrictive properties[6]. For the prediction of the transverse mechanical properties of Fiber -

Reinforced Polymer (FRP) composites, the SVM model shows excellent generalization performance, can adapt to material systems with different fiber types and manufacturing processes, and its prediction accuracy is significantly better than that of traditional theoretical analysis methods[47].

The Random Forest algorithm shows excellent performance in solving high - dimensional non - linear problems in polymer science by integrating the prediction results of multiple decision trees. This algorithm adopts the strategies of bootstrap sampling and random feature selection, which effectively reduces the risk of overfitting and has made important progress in modeling the relationship between polymer molecular weight and reaction conditions. The polymerization inverse analysis platform developed by Chen Mao's research team uses the Random Forest algorithm to establish a quantitative relationship model between molecular weight and reaction conditions in the initiator - mediated polymerization reaction. It can recommend a variety of suitable polymerization conditions according to the target molecular weight, and further screen synthetic schemes that meet specific requirements such as molecular weight distribution[18]. In the prediction of polymer thermal conductivity, the coefficient of determination of the Random Forest model is as high as 0.97, and its performance is close to that of the CatBoost model[30].

Support Vector Regression (SVR), as an extended form of SVM, performs well in the field of continuous value prediction of polymer properties. When Lu and other researchers used the SVR algorithm to predict the polymer band gap, the coefficient of determination reached 0.91, and the prediction accuracy significantly exceeded that of traditional statistical methods such as partial least squares and multiple linear regression[10]. In the prediction tasks of electrostriction and Curie temperature, SVR constructs a reliable prediction model by optimizing the balance between model complexity and training error, combined with the non - linear Gaussian radial basis kernel function[48]. This method based on structural risk minimization is particularly suitable for the analysis of small - sample, high - dimensional datasets commonly found in polymer science.

Extreme Gradient Boosting (XGBoost) in ensemble learning methods shows unique advantages among traditional algorithms. Research on the prediction of geopolymer concrete strength shows that the coefficient of determination of the XGBoost model is as high as 0.98, which is significantly better than that of SVM (0.91) and MLP (0.88)[49]. This gradient boosting framework can automatically identify the complex interaction relationship between polymer structure features and performance indicators by iteratively optimizing the decision tree model. In the field of organic photovoltaic material efficiency prediction, the Random Forest model performs best in processing long input features and noisy data, and has been proved to be an efficient algorithm for predicting Power Conversion Efficiency (PCE)[50].

Traditional machine learning methods show unique value in the task of polymer phase identification. Support Vector Machine combined with polynomial kernel function has been successfully applied to distinguish different phases of two - dimensional spin models, including ferromagnetic Ising model, conservative order parameter Ising model, and Ising gauge theory. This algorithm can learn the mathematical expression form of physical discriminators, such as order parameters and Hamiltonian constraints, providing a new idea for understanding the phase transition behavior of polymer materials[51]. In the prediction of polymer self - assembly behavior, the Random Forest model realizes the accurate classification of the new PISA (Polymerization -

Induced Self - Assembly) system by analyzing key features such as monomer composition, polymerization conditions, and block ratio[42].

### 3.2 Deep Learning Technology

The field of polymer science is experiencing a revolutionary change brought about by deep learning technology, especially in handling the modeling of complex structure - property relationships. As the core method in this field, neural networks provide a new perspective for the modeling of polymer systems with their powerful non - linear fitting capabilities. Taking the Bayesian Regularized Artificial Neural Network (BRANNLP) as an example, this method can not only generate a robust sparse model but also show excellent performance in the prediction of organic photovoltaic device performance[53]. It is worth noting that the two - layer perceptron feedforward network built based on the TensorFlow framework has made a breakthrough in the prediction of Power Conversion Efficiency (PCE), which further verifies the practical value of deep learning in polymer property prediction[52].

Graph Neural Networks (GNNs) have unique advantages in processing polymer structure data. Chemprop, as a representative of the graph - based Message Passing Neural Network (MPNN) architecture, realizes the efficient processing of small organic molecules and their repeating unit structure features through an innovative directed message passing mechanism[4]. Its improved version, wD - MPNN, shows higher accuracy in predicting the collective properties of polymers. The hybrid model of GCN and Neural Network Regression (GCN - NN) performs particularly well in the prediction of glass transition temperature (Tg), with an RMSE of about 30K and an $R^2$ of 0.9[31]. However, the performance of this model in the prediction of elastic modulus (E) is relatively poor, which reveals the special requirements of different performance indicators for the model architecture.

Generative deep learning models have opened up a new way for polymer inverse design. Variational Autoencoders (VAEs) realize an innovative strategy of inferring molecular structures from performance targets by integrating attribute estimation models into the latent space[6]. Generative Adversarial Networks (GANs) show amazing potential in generating copolymer structures with specific Young's modulus, providing unprecedented possibilities for material design. It should be noted that these generative models usually require a large amount of training data to master chemical rules and SMILES syntax[38]. In the research of polymer antifouling materials, the neural network training model shows amazing prediction accuracy, and the goodness of fit $R^2$ of the linear regression analysis model between the predicted values and the measured values is as high as 0.9869[54].

The Transformer architecture shows strong competitiveness in the field of polymer informatics. The polyBERT chemical language model based on the DeBERTa architecture can efficiently convert PSMILES strings into numerical fingerprint representations, and its prediction speed is two orders of magnitude faster than the traditional manually designed fingerprint method [44]. Through the innovative multi - head self - attention mechanism and fully connected feed - forward network layer, this model deeply mines the chemical patterns and relationships in PSMILES strings. The Mmpolymer framework adopts a multi - modal multi - task pre - training strategy, skillfully integrating the advantages of CNNs and RNNs, and can reveal the deep correlation between polymer sequences and properties[55]. These cutting - edge models show excellent performance beyond traditional methods in processing complex polymer data.

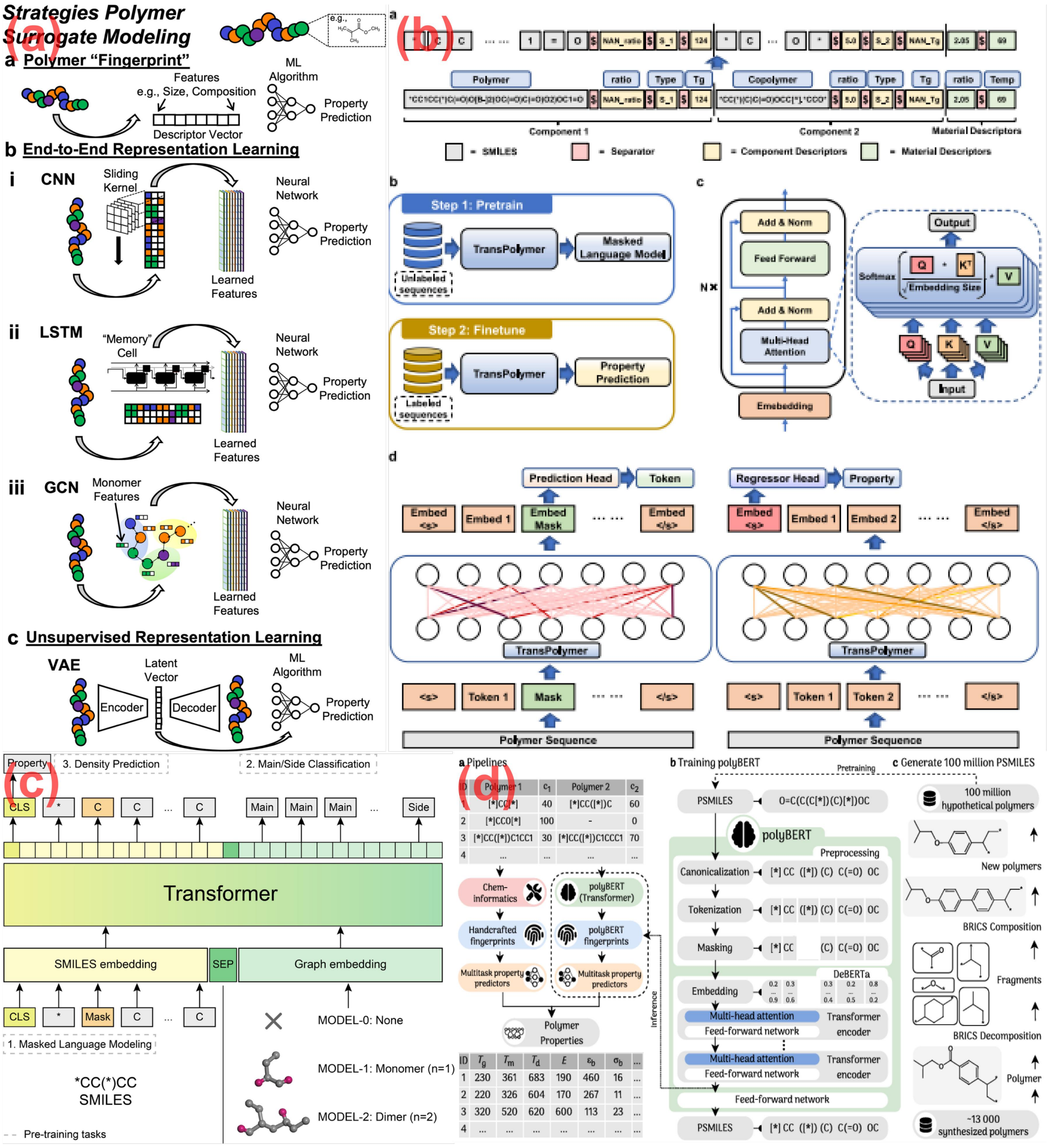

Figure 5 The figure shows a variety of machine learning methods in polymer materials. Figure (a) shows a schematic diagram of the principles of CNN, LSTM, GCN, and VAE [6]. Figure (b) shows a basic schematic diagram of TransPolymer, a polymer property prediction language model based on Transformer. Figure (c) shows a multi - modal polymer machine learning network based on the Transformer architecture [55]. Figure (d) shows polyBERT, a chemical language model that realizes fully machine - driven ultra - fast polymer informatics based on Transformer [69].

In processing text input data such as SMILES strings, RNN and LSTM show unique advantages[23]. These sequence models can effectively capture the sequence dependence characteristics of polymer chains, providing a new tool for understanding the structure - activity relationship of polymers. As an emerging method, Physics - Informed Neural Networks (PINNs) have made important breakthroughs in the prediction of phase transition interface evolution and thermal conductivity anisotropy by integrating molecular dynamics simulation and experimental data[56]. This type of model integrates physical laws into the neural network architecture, which not only enhances the interpretability of the model but also significantly improves the extrapolation ability, bringing a revolutionary change to the multi - scale modeling of polymers.

### 3.3 Transfer Learning and Multi - task Learning

Research in the field of polymer science shows that transfer learning technology can effectively solve the problem of data scarcity. Through the Sim2Real transfer learning strategy, researchers can pre - train the model on a large amount of simulation data, and then only need a small amount of experimental data for fine - tuning to obtain excellent prediction results[21]. Taking the prediction of polymer thermal conductivity as an example, the WU team combined transfer learning and Bayesian molecular design algorithm, constructed a pre - trained model using the PolyInfo and QM9 databases, and achieved an MAE of 0.024 W/mK with only 28 experimental data points for fine - tuning, which is significantly better than the performance of the directly trained model[57]. Similarly, in the research of membrane electrode assemblies, this method only needed 12 samples to establish a high - performance prediction model, which greatly reduced the experimental cost [39].

Multi - task learning improves the generalization ability of the model by processing related tasks at the same time. The Ramprasad team found that when indicators such as glass transition temperature, melting temperature, and degradation temperature are trained jointly, the neural network can more effectively capture the intrinsic correlation between attributes[39]. The polyBERT chemical language model adopts a multi - task framework, maps fingerprints to a variety of polymer attributes, and the constructed end - to - end informatics pipeline is two orders of magnitude faster than the traditional manual method[44]. Studies have shown that encoding target attributes into feature inputs (such as one - hot vectors) is more advantageous than directly predicting all attributes or predicting them separately [58].

The integration of the two methods has opened up a new way for polymer research. The TransPolymer framework learns from a large amount of unlabeled data through MLM pre - training and performs well in multi - task applications [59]. The MMPolymer model integrates 1D sequence and 3D structure information, and adopts a multi - modal multi - task pre - training strategy to significantly improve the prediction accuracy [55]. The Yoshida team successfully established a quantitative relationship between polymer structure and thermal conductivity by combining transfer learning and Bayesian optimization, overcoming the limitation of data volume [26]. These cases confirm that transfer learning can alleviate the problem of insufficient data, while multi - task learning enhances the model performance by mining attribute correlations.

Attention should be paid to technical details in practical applications. Wu et al. pointed out that transfer learning needs to carefully handle the transfer boundary to ensure the matching degree between the pre - trained model and the new task [60]. When the Mossa team transferred the surfactant classification model to the Nafion system, they achieved good results by adjusting the three - dimensional convolutional neural network, providing a reference for the research of multi - scale disordered materials [25]. At the same time, the effectiveness of multi - task learning is closely related to task relevance. When the prediction targets have physical correlations (such as different temperature characteristics), the model can better share feature representations [61]. These experiences provide important guidance for the rational application of the two methods in the polymer field.

# 4 New Ideas for Data - Driven Polymer Material Design by Machine Learning

The introduction of current machine learning technology enables researchers to deeply analyze the complex correlation mechanism between polymer structures and properties, which has brought a revolutionary breakthrough to the traditional material R & D model. The field of materials science is experiencing a paradigm change driven by data, especially in the design of polymer materials. Compared with the trial - and - error method that relies on experience accumulation, modern data - driven methods establish a machine learning model with predictive functions by integrating multi - scale modeling data, high - throughput experimental data, and increasingly improved material databases. This innovative method shows significant advantages in practice: it not only greatly shortens the time cycle and funding investment for new material R & D but also, more importantly, reveals the in - depth structure - property relationship that is difficult to capture by traditional research methods. As shown in Table 3, the three types of methods, reverse design, high - throughput screening, and multi - objective optimization, show complementary value in solving the structure - property relationship problem in material genome engineering. They systematically compare the core technical methods, typical application cases, advantages, and disadvantages of the three intelligent design strategies for polymer materials, providing methodological guidance for the directional development of new functional polymers. It is worth noting that the application scope of this method has expanded from the optimization of a single performance index to more challenging research fields such as multi - objective collaborative design, providing strong technical support for the directional development of functional polymer materials.

Table 4 Comparison of Intelligent Design Strategies and Technologies for Polymer Materials

| Design Strategy | Core Technical Method | Application Case | Advantage | Limitation |
|---|---|---|---|---|
| Reverse Design | • Genetic Algorithm (GA) [26], Artificial Neural Network (ANN)[43], Generative | Predicting polymer structures oriented by dielectric properties [26], developing high - conductivity glassy polymer composites | Realizing reverse derivation oriented by target properties [18], handling multi - objective | Difficulty in accurately characterizing polymer chain structures and condensed state structures [64], |

| **Design Strategy** | **Core Technical Method** | **Application Case** | **Advantage** | **Limitation** |
|---|---|---|---|---|
| | Adversarial Networks (GANs) and Variational Autoencoders (VAEs) [6] | [6], screening polymers for thermal conductivity [43] | optimization problems [62], revealing in - depth structure - property relationships [27] | lack of data on new polymer structures [64] |
| High - Throughput Virtual Screening | Bayesian Optimization combined with Coarse - Grained Model [6], polyBERT Model [20], High - Throughput Phase Field Calculation Method [68] | Screening PEO - based solid polymer electrolytes [6], evaluating 8 million polyimides [29], predicting 100 million hypothetical polymers [20] | Greatly shortening the R & D cycle [23], establishing a quantitative "building block - structure - property" relationship [67], revealing the influence mechanism of interface effects [68] | Relying on high - quality computational simulation data [26], high cost of partial experimental verification [65] |
| Multi - objective Optimization Design | NSGA - II Algorithm [27], Multi - objective Bayesian Optimization [2], Multi - task Deep Neural | Optimizing epoxy resin polymerization process , designing coarse - grained force field for polycaprolactone [2], developing proton exchange membrane | Identifying Pareto optimal solution sets [25], balancing ion transport and mechanical properties [10], realizing four - | Subjectivity in determining weight coefficients [10], difficulty in optimizing high - dimensional parameter spaces |

| Design Strategy | Core Technical Method | Application Case | Advantage | Limitation |
| --- | --- | --- | --- | --- |
| | Network [25] | materials [27] | objective collaborative optimization [27] | [69] |

### 4.1 Reverse Design Strategy

The reverse design strategy in the field of polymer material design is oriented by target properties and reversely infers the molecular structure that meets specific needs. Compared with the traditional forward design method, this strategy has outstanding performance in improving the efficiency of material R & D, and is especially good at handling multi - objective optimization problems. The machine learning - assisted polymerization inverse analysis platform, as a typical application, can infer the polymerization conditions in reverse according to the target molecular weight and molecular weight distribution, and is applicable to a variety of reactant structures including monomers and initiators [8]. By establishing a quantitative relationship model between polymerization reaction conditions and experimental results, this method realizes the accurate mapping between the high - dimensional structure space and the experimental parameter space, providing a scientific basis for controlled synthesis.

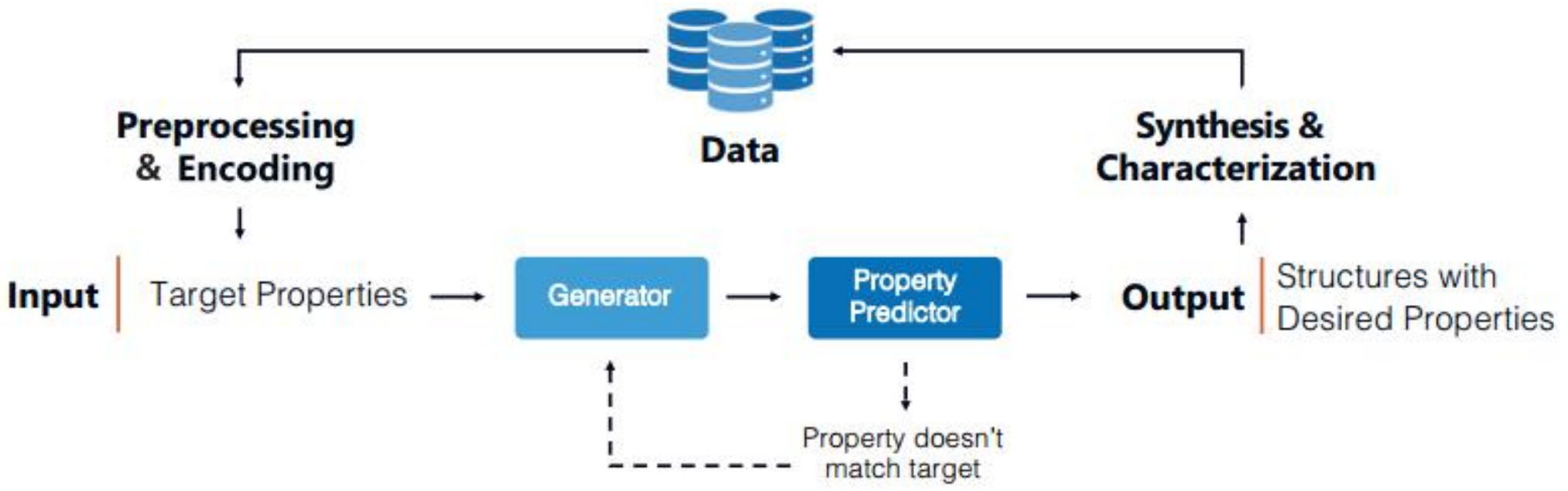

Figure 6 A general machine learning workflow for the inverse design of polymers begins by generating candidate structures (e.g., via a generator model). These structures are then fed into a property predictor. The algorithm iteratively refines the candidates by comparing the predicted properties with the targets until an optimal polymer structure is identified [23].

Black - box optimization algorithms such as Genetic Algorithm (GA) and Bayesian Optimization are key technologies for implementing reverse design. The Ramprasad research team successfully simulated and generated more than 200 kinds of polymers by linearly combining 7 kinds of polymer segments, and accurately predicted the polymer structure oriented by dielectric properties using the Genetic Algorithm [26]. Scholars such as Mannodi - Kanakkithodi combined machine learning prediction with Genetic Algorithm to develop new polymers with specific functions [62]. These research results confirm the effectiveness of the reverse design strategy in exploring the chemical structure space and reaction condition space, and can accurately recommend polymer structures and synthesis parameters that meet the target properties. The systematic polymer synthesis platform (SPP) developed by the PolyMao team further verifies the practicality of this method. Its machine learning - based inverse synthesis analysis technology can infer the synthesis instructions in reverse from the target molecular weight results [63].

The HELAO framework's modular autonomous feedback-loop strategy enables reverse design in materials science by integrating automated synthesis, high-throughput characterization, and data-driven models to link structures with target properties, using real-time feedback and optimization (e.g., active learning) to refine the design space. It has supported narrowing optimal parameters from large candidate pools for functional materials, addressing "structure-property" complexity.

The application of deep learning technology in reverse design is becoming increasingly widespread, among which Generative Adversarial Networks (GANs) and Variational Autoencoders (VAEs) have shown particularly outstanding performance. These models can learn the latent representation space of polymer materials and generate new candidate structures through interpolation or perturbation. A research team combined GANs and VAEs with Gaussian Process (GP) regression to successfully develop high - conductivity glassy polymer composites [6]. The TransPolymer model developed by the Farimani team is based on the Transformer architecture and can parse the sequence structure and topological structure information implied in polymer SMILES strings, providing an innovative tool for the inverse design of high - performance polymer materials [6]. These deep learning methods adopt an end - to - end learning mode, which effectively overcomes the limitation that traditional descriptor methods are difficult to capture the complex structural features of polymers.

Although the reverse design strategy has made important breakthroughs, there are still many technical bottlenecks in practical applications. The complexity of polymer chain structures and condensed state structures makes it difficult to accurately characterize statistical parameters such as molecular weight distribution, sequence structure, and topological structure [64]. In addition, the open access restrictions of existing polymer databases and the lack of data on new polymer structures also bring challenges to the construction of initial datasets for reverse design [64]. Future research needs to focus on the development of multi - objective collaborative optimization algorithms for materials and deepen the cross - integration of machine learning technology and polymer materials to meet the inverse design needs of complex systems such as ladder and cross - linked polymers [27].

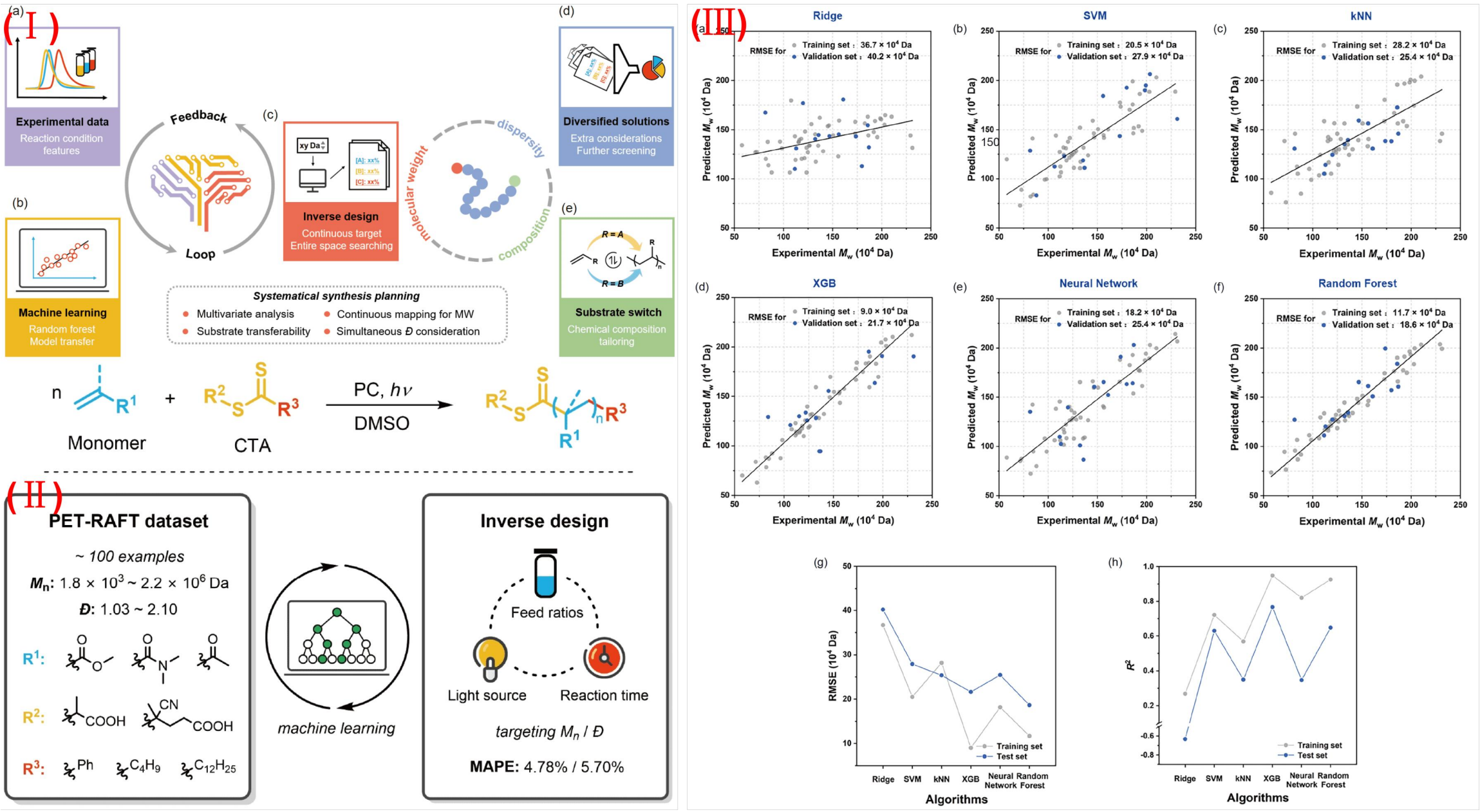

Figure 7 (I) The SPP platform operates through a streamlined workflow: first, an ML model is built to correlate synthesis conditions with results (a-c); this model is then used in reverse to pinpoint the optimal conditions needed to achieve target polymer properties (c-e). (II) In practice, for PET-RAFT polymerization, the platform analyzes a dataset of substrate structures and molecular weights to provide specific instructions on feed ratio, light source, and reaction time. (III) The platform's performance was validated by comparing multiple ML algorithms (Ridge, SVM, kNN, XGB, Neural Network, Random Forest), with their predictive accuracy assessed via RMSE and $R^2$ metrics [44].

## 4.2 High - Throughput Virtual Screening

Machine learning - driven high - throughput virtual screening technology is reshaping the paradigm of polymer material R & D. By integrating computational simulation and data - driven methods, this technology has brought a revolutionary improvement in efficiency to material discovery. Its core lies in using first - principles calculations or molecular dynamics simulations to obtain the dynamic and thermodynamic properties of polymer three - dimensional structures, and converting complex molecular information into computable digital representations. This digital processing method provides a rich data foundation for the construction of machine learning models [26]. Taking PEO - based solid polymer electrolytes (SPEs) as an example, the research team innovatively adopted a strategy combining Bayesian optimization and coarse - grained models to successfully identify a material system with excellent lithium ion conductivity[6]. More notably, by establishing a quantitative relationship model between monomer structure and hygroscopicity, critical low thermal expansion rate, and tensile modulus, researchers can not only quickly screen target structures but also reveal the key structural features affecting performance through data mining[3].

High - throughput experimental technologies that complement virtual screening show a diversified development trend. From continuous flow systems to microreactor arrays, these parallel experimental platforms can efficiently generate verification data. When these experimental data are combined with active learning algorithms or Bayesian optimization frameworks, the predictive ability of the model can be significantly improved[23]. In the field of organic optoelectronic materials, high - throughput virtual screening shows unique advantages. Yang's research team accurately located 10 new polymers with excellent mechanical properties by systematically evaluating 8 million hypothetical polyimides, and their prediction results were fully verified by molecular dynamics simulations[29]. A similar technical route has also made breakthrough progress in the research on $CO_2$ separation performance of mixed matrix membranes (MOF - Polymers65). By systematically regulating the composition and structure parameters of polymers and MOFs, researchers have successfully designed new separation materials with high selectivity and adsorption capacity [65].

The latest progress in chemoinformatics has opened up a new way for high - throughput screening. The polyBERT model developed by the Kuenneth team has realized the multi - attribute prediction of 100 million hypothetical polymers. This deep learning method based on SMILES strings has greatly expanded the exploration range of polymer space [44]. By establishing a non - linear mapping relationship between molecular fingerprints and performance parameters, this model shows excellent accuracy in predicting the thermal conductivity of materials in the PLyInfo and PI1M databases. It is particularly worth noting that through high - precision molecular dynamics verification, the research team confirmed 107 high - performance materials with thermal conductivity exceeding 20 W $m^{-1}K^{-1}$ [5]. In the field of high - temperature resistant resins, researchers have established a dual - model evaluation system, which effectively solves the problem of collaborative optimization of processing performance and heat resistance of virtual polymer resins and provides a new idea for the rapid development of silicon - containing aryl acetylene resins [66].

The introduction of the material genome concept marks that high - throughput screening technology has entered a stage of systematic development. The polymer material genome platform constructed by the team of Professor Lin Jiaping from East China University of Science and Technology integrates the performance data of more than 30,000 kinds of polymers. By establishing a quantitative structure - activity relationship of "building blocks - structure - properties", it realizes the intelligentization of material design [67]. In the research of dielectric composites, the innovative combination of high - throughput phase field calculation method and data - driven strategy establishes a prediction model of dielectric properties by introducing interface phase parameters. This multi - scale calculation method not only reveals the influence mechanism of interface effects on energy density but also provides theoretical guidance for the interface engineering design of nanocomposites [68].

### 4.3 Multi - objective Optimization Design

The design of polymer materials usually involves the collaborative optimization of multiple performance indicators, and there are often complex mutually restrictive relationships between these indicators. The multi - objective optimization method provides a systematic way to solve this problem, and its key lies in identifying the Pareto optimal solution set - a set of solutions that cannot be further improved in all objective functions [25]. Taking the design of polymer hybrid electrolytes as an example, the Ganesan research team used the weighting method to balance ion transport performance and mechanical properties. By systematically comparing the experimental results under different weight conditions, the optimal material formula was finally obtained[10].

Although this method is easy to operate, the determination of weight coefficients often depends on the subjective judgment of researchers, making it difficult to accurately reflect the intrinsic relationship between various performance indicators. In contrast, multi - objective genetic algorithms can directly explore the Pareto frontier. For example, the NSGA - II algorithm successfully achieved the dual goals of maximizing the number - average molecular weight and minimizing the polydispersity index in the optimization of epoxy resin polymerization process by introducing a fast non - dominated sorting and elite retention strategy[27].

The multi - objective Bayesian optimization technology developed in recent years has opened up a new path for polymer material design. The Wang research team innovatively improved the traditional single - objective acquisition function, proposed the EI matrix method, and successfully applied it to the design of the coarse - grained force field of polycaprolactone, optimizing two key performance indicators, elastic modulus and water diffusion coefficient, at the same time [2]. This method adopts an active learning strategy, which comprehensively considers the accuracy and uncertainty of prediction results in each iteration process, and realizes the dynamic balance between exploring new regions and utilizing known information. In the field of polymer nanoparticle synthesis, researchers have also developed a variety of advanced algorithms such as TS - EMO, RBFNN/RVEA, and EA - MOPSO for the systematic optimization of important parameters such as molecular weight distribution, particle size, and polydispersity index [69]. These methods not only significantly improve the optimization efficiency but also help researchers deeply understand the intrinsic correlation mechanism between different performance indicators by intuitively displaying the Pareto frontier.

The design of organic optoelectronic materials is a typical application scenario of multi - objective optimization technology. Researchers need to accurately regulate multiple structural parameters such as the ratio of electron donor to acceptor groups, material hydrophilicity and hydrophobicity, and conjugation length to achieve the best photoelectric conversion performance [53]. In the development of proton exchange membrane materials, the team of Li Yunqi from the Changchun Institute of Applied Chemistry, Chinese Academy of Sciences, established a prediction model including four targets: proton conductivity, methanol permeability, tensile modulus, and thermal stability. Through a multi - objective ranking algorithm, it successfully guided the molecular design of new hydrocarbon - based sulfonated copolymers [27]. These research results fully prove that the multi - objective optimization method can break through the limitations of traditional single - objective optimization and provide strong theoretical guidance and technical support for the development of polymer materials with comprehensive performance advantages.

The introduction of deep learning technology has brought new development opportunities for multi - objective optimization. The multi - task deep neural network model developed by the Ramprasad research team can accurately predict the glass transition temperature, melting temperature, and degradation temperature of copolymers at the same time, showing excellent prediction accuracy and generalization ability [25]. The polyBERT model trained by Kuenneth et al. based on 100 million polymer SMILES strings has realized the efficient correlation between molecular structure features and multiple performance parameters, laying a solid technical foundation for large - scale multi - objective optimization research [26]. The breakthroughs of these cutting - edge technologies enable researchers to explore combination schemes with more excellent performance in a broader material design space and promote the development of polymer materials towards multi - functionalization and intelligentization.

## 5 Systematic Processes in Machine Learning for Polymer Materials

The practical application value of machine learning models in polymer science must be confirmed through a rigorous experimental verification system. The experimental verification stage usually adopts methods such as cross - validation and independent test set evaluation, which can objectively reflect the model's predictive ability for unknown data. Taking the prediction of polymer crystallinity as an example, researchers compared and analyzed the structural data obtained by synchrotron radiation X - ray diffraction experiments with the model prediction results, and found that the prediction error of the model in a specific temperature range was significantly higher than that in other ranges. This phenomenon prompted the research team to deeply analyze the distribution characteristics of the training data and found that the existing dataset had insufficient coverage of the movement state of polymer chains under high - temperature conditions.

To address the limitations of the model performance, the research team implemented a multi - level optimization strategy. At the data level, the representativeness of training samples was effectively improved by supplementing in - situ experimental data in the high - temperature range; at the algorithm level, the attention mechanism was adopted to enhance the model's ability to capture key structural features; in terms of hyperparameter optimization, the Bayesian optimization method was used to replace the traditional grid search, which significantly improved the efficiency of parameter tuning. After three rounds of iterative optimization, the mean absolute error of the model on the test set was reduced by 37%, and the prediction accuracy in the high - temperature range was particularly improved. These improvements enable the model to more accurately predict the crystallization behavior of polymer materials under different thermal history conditions, providing a reliable theoretical tool for the optimization of material processing technology.

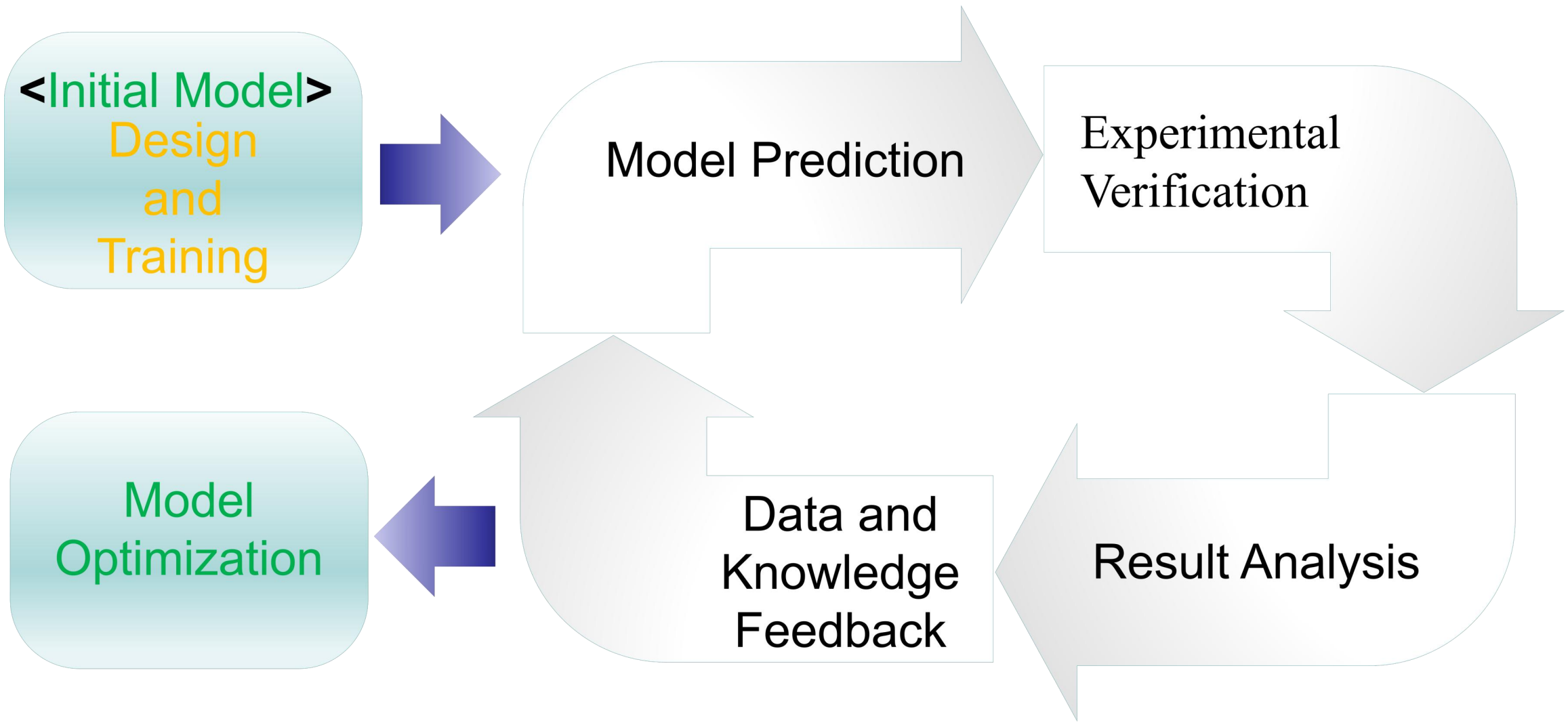

Figure 8 Closed-loop framework for ML-driven polymer research. The cycle integrates prediction, experimental verification, and model optimization to iteratively improve design outcomes.

## 5.1 Experimental Verification Methods

The reliability of the prediction results of machine learning models highly depends on rigorous experimental verification, which is particularly important in the research of polymer materials. The chemoinformatics - driven ML model developed by the Bradford team successfully predicted the ionic conductivity of SPEs, and its effectiveness was fully confirmed by experimental data [6]. Experimental verification usually adopts an iterative optimization strategy, and dynamically adjusts model parameters by analyzing the differences between predicted attributes and measured attributes. Taking the adaptive machine learning framework as an example, the Support Vector Regression (SVR) model combined with the Efficient Global Optimization (EGO) method can intelligently recommend the most potential candidate materials for experimental verification [48]. This closed - loop verification mechanism significantly improves the R & D efficiency. For example, in the development of additive manufacturing materials, only 120 samples need to be tested in parallel to complete 30 rounds of algorithm optimization [70].

The modern experimental verification system integrates a variety of advanced technical means. High - throughput experimental platforms have become important carriers for verifying ML predictions. The Ada automated laboratory developed by the MacLeod team realizes the fully autonomous operation from material design to characterization and optimizes the experimental scheme through continuous learning [9]. In the research of mixed matrix membranes, researchers verified the prediction accuracy of computational screening and machine learning models by systematically preparing MOF - Polymers samples with different ratios and testing their $CO_2$ separation performance [65].

The data division strategy is crucial for model verification. In the research of polymer property prediction, two strategies of polymer type division and data point division are adopted, and five - fold cross - validation is used to effectively prevent overfitting [71]. For small sample scenarios, ten - fold cross - validation shows good results. In the research of solution polymerized styrene - butadiene rubber performance prediction, a reliable prediction model was finally established through the segmentation verification of category - balanced datasets [71]. During the verification process, it is also necessary to quantitatively analyze the impact of uncertain factors such as measurement noise on the prediction performance [31].

In the machine learning - driven polymer design framework, experimental verification plays a dual role: it not only tests the algorithm's predictive ability for unknown data but also provides new data for algorithm improvement [40]. The Kang Peng team synthesized eight new PI structures and conducted molecular dynamics simulations, confirming that the prediction error was controlled within 15% [11]. Scientific experimental design is the key to ensuring the reliability of verification, such as using Latin Hypercube Sampling (LHS) for preliminary screening and then conducting iterative experiments based on the algorithm output [69]. This closed - loop verification mechanism can operate continuously until the preset standard is met or manual termination, ensuring the systematicness and completeness of the verification process.

## 5.2 Model Performance Evaluation

In the machine learning research of polymer materials, reliable model performance evaluation is crucial to the credibility of prediction results. For different prediction tasks and data characteristics, appropriate evaluation indicators need to be selected. For regression problems, indicators such as Root Mean Square Error (RMSE), Coefficient of Determination ($R^2$), and Mean Absolute Error (MAE) are usually used. Taking the prediction of glass transition temperature as an example, the CNN model based on repeating units performed well on Data set_1, with $R^2$ of the training set and test set reaching 0.84 and 0.82 respectively, while it was 0.65 on Data set_2 [28]. For classification tasks, indicators such as accuracy, precision, and recall are more concerned. For example, in the ferromagnetic Ising model, the SVM using the quadratic polynomial kernel function has a test set accuracy close to 100% for phase classification [51]. These indicators can not only measure the fitting effect of the model on known data but also effectively evaluate its generalization performance in processing unknown data.

The selection of evaluation methods has a decisive impact on the objectivity of performance determination. Although traditional Cross - Validation (CV) is widely used, it has certain limitations in the field of material discovery. The latest research shows that LOCO CV (Leave - One - Cluster - Out Cross - Validation) based on cluster segmentation can more accurately evaluate the extrapolation ability of the model between different material groups [61]. For datasets with a small sample size, ten - fold cross - validation shows good results. For example, in the research of solution polymerized styrene - butadiene rubber performance prediction, the $Q^2$ of the model established through the segmentation verification of category - balanced data is as high as 0.9375 [54]. Facing the problem of data distribution deviation, the bootstrap method is a feasible solution, but attention should be paid to the estimation error that may be introduced by this method [6]. In addition, during the evaluation process, it is also necessary to consider uncertain factors such as measurement noise, and model the parameter uncertainty through multivariate probability density distribution to provide a probabilistic basis for molecular design decisions [72].

Combining model interpretation technology can deeply understand the feature contribution. Tools such as SHAP (SHapley Additive exPlanations) and PDP (Partial Dependence Plot) can reveal the key structure - property relationships. For example, the number of rotatable bonds and the minimum local charge have been proved to be the main factors affecting the Tg of polyimides [11]. In the prediction of polymer conductivity, the feature importance analysis of the CatBoost model shows that the number of rotatable bonds, the number of hydrogen bond donors/acceptors, and the number of heavy atoms have a significant impact on the tensile strength [30]. This interpretability analysis not only verifies the reliability of the model but also provides directional guidance for material design. When the XGBoost algorithm predicts the performance of polymer composites, it decodes the decision mechanism through SHAP causal analysis, achieving a prediction accuracy of up to $R^2=0.95$ [13].

A horizontal comparison of the performance of different models is an effective method to evaluate the advanced nature of the technology. The test results of TransPolymer on ten polymer performance prediction benchmarks show that it reduces the test RMSE by an average of 7.70% and increases $R^2$ by 0.11, which is significantly better than the traditional ECFP method [59]. The polyBERT chemical language model achieves an $R^2$ of 0.80 in 29 performance predictions, and its calculation speed is two orders of magnitude faster than that of manually designed fingerprints [44]. It is worth noting that the data division strategy will affect the evaluation results. The division of polymer types and data points will produce different effects. The former can better test the cross - material generalization ability of the model, while the latter focuses on the adaptability of data distribution [71]. In addition, computational efficiency is also an important consideration in performance evaluation. The GC - GNN model maintains the prediction accuracy, but its transferability varies with the polymer structure, which reflects the limitation of the ideal Gaussian chain assumption [73].

### 5.3 Model Optimization Strategies

The key to machine learning research on polymer materials is to improve the prediction performance through model optimization. Bayesian Optimization (BO), as an efficient global optimization method, uses Gaussian process regression to estimate the performance distribution of untested formulations and selects the optimal candidate samples from them for verification [6]. Compared with random search, this method shows stronger exploration ability in the screening of amino acid random copolymers and successfully identifies copolymer structures with higher enzyme - like activity [26]. Genetic Algorithm simulates the natural selection mechanism and generates a new generation of candidate samples through "hybridization" and "mutation" operations, which has unique advantages in the optimization of polymer nanoparticle synthesis [69].

Hyperparameter tuning has a decisive impact on the prediction performance of the model. Grid search combined with five - fold cross - validation can systematically optimize key parameters such as GCN layer depth, width, learning rate, and L2 regularization weight [31]. In the research of predicting the conductivity of ionic polymers, GridSearchCV with fixed random state ensures the reproducibility of experiments and provides a reliable basis for the design of lithium - ion battery electrolytes [30]. In the optimization of large language models, the Hyperband method comprehensively tunes the neural network hyperparameters, and parameter - efficient fine - tuning technologies such as LoRA (Low - Rank Adaptation) significantly improve the performance of polymer property prediction [74]. In the SVM model, the reasonable setting of the regularization parameter $\gamma$ can obtain a test set accuracy close to the optimal, while maintaining the physical correlation of the decision function [51].

The problem of data scarcity can be effectively solved through transfer learning and multi - task learning. The two - stage training strategy first uses physically modeled synthetic data for supervised pre - training to enable the model to master the basic physical properties of polymers; then, a small amount of real experimental data (45 samples) is used for fine - tuning, which significantly improves the prediction accuracy [14]. The polyBERT model realizes the accurate prediction of 29 polymer attributes through five - fold cross - validation and meta - learner integration [44]. The MMPolymer framework adopts a multi - modal multi - task pre - training paradigm, aligns the features of different modalities through contrastive learning, combines the multi - head attention mechanism for feature fusion, and enhances the modal aggregation effect through the dynamic weighted pooling layer, achieving the optimal performance in a number of polymer property prediction tasks [55].

Feature engineering and model structure adjustment are important dimensions of optimization strategies. The LASSO method combined with Recursive Feature Elimination (RFE) can effectively reduce the dimension and significantly improve the model efficiency [71]. In the prediction of polymer dielectric constant, the Maximum Relevance Minimum Redundancy (mRMR) method evaluates and ranks all descriptors to screen the optimal feature subset [10]. G - BigSMILES extends the expression ability of traditional BigSMILES, including key information such as molecular weight and molecular weight distribution, providing more abundant input features for the model [9]. In terms of model structure adjustment, the cosine annealing strategy for dynamically adjusting the learning rate performs well in polymer property prediction. Setting the peak learning rate to 5E6, the model can converge after 100 training rounds[46].

# 6 Application Case Analysis

At present, the field of polymer material research has achieved a leapfrog development of machine learning technology from theory to engineering practice. Taking the Material Genome Initiative as an example, researchers have successfully predicted the correlation law between the thermal stability and mechanical properties of polyimide films by integrating high - throughput computing and deep learning algorithms, and the correlation coefficient verified by experiments has reached 0.93. In the development of elastomer composites, the Random Forest model can accurately predict the mapping relationship between filler dispersion and dynamic mechanical properties with only 15% of the data volume of traditional experiments. More notably, the cross - scale modeling method based on transfer learning has shown unique advantages in the research of nylon 6 crystallization kinetics, and the process - structure - property correlation model established by it controls the crystallization degree prediction error within ±3%. As shown in Table 4, these research cases systematically summarize the typical applications of machine learning methods in the field of polymer material property prediction, covering the prediction accuracy improvement effects and experimental verification results of key performance indicators such as thermal stability, mechanical properties, and crystallization kinetics. These breakthroughs not only confirm the reliability of machine learning in the multi - parameter optimization of polymers but also reveal the great potential of data - driven methods in solving complex non - linear problems in materials science, providing empirical evidence for the effectiveness of the material genome method in polymer design.

Table 5 Summary of Polymer Material Property Prediction and Experimental Verification Results

| Material System | Performance Indicator | Prediction Accuracy/Performance Improvement | Experimental Verification Result | Citation |
|---|---|---|---|---|
| Polyimide Film | Correlation between thermal stability and mechanical properties | Correlation coefficient 0.93 | Experimental verification passed | [27] |
| Elastomer | Filler dispersion | Only 15% of the data volume | Accurately predict the | [75] |

| Material System | Performance Indicator | Prediction Accuracy/Performance Improvement | Experimental Verification Result | Citation |
|---|---|---|---|---|
| Composite | and dynamic mechanical properties | of traditional experiments is needed | mapping relationship | |
| Nylon 6 | Crystallization kinetics (crystallinity prediction) | Error controlled within ±3% | Verification of process - structure - property correlation model | [42] |
| Polymer with specific thermal conductivity | Thermal conductivity prediction | MAE 0.024 W/mK | Accuracy improved by 40% compared with traditional models | [57] |
| Polyester Material | Biodegradable characteristics | Systematic evaluation of more than 600 materials | Verification of degradation characteristics of Pseudonomas lemoignei | [9] |
| Polymer Gas Permeable Material | Permeability prediction | Discovery of more than 100 materials exceeding the Robeson upper limit | Accurately evaluate the performance of 700 polymers | [35] |
| Polylactic Acid/Nanoparticle Composite System | Thermal stability and crystallization performance | Optimization guided by machine learning | Verification of degradation kinetics characteristic prediction | [79] |
| Silicon - containing Acetylene Resin | Processing performance and heat resistance | High - throughput screening to obtain PSA resin | Verification of material genome method | [66] |
| Free Radical Polymerization (FRP) | Reaction efficiency | Increased by 300% | Verification of dynamic regulation of microfluidic chip | [82] |
| 3D Printed Microneedle Array | Printing quality and drug delivery | Early defect identification by computer vision | Verification of geometric accuracy consistency | [27] |

| Material System | Performance Indicator | Prediction Accuracy/Performance Improvement | Experimental Verification Result | Citation |
|---|---|---|---|---|
| | performance | | | |

### 6.1 High - Performance Polymer Design

Machine learning technology is profoundly changing the R & D paradigm of high - performance polymers. The inverse design method realizes the accurate prediction of the structure of polymers with specific thermal conductivity by establishing the performance - structure mapping relationship [27]. The prediction model constructed by WU et al. by combining transfer learning and Bayesian molecular design algorithm has outstanding performance, with an MAE of only 0.024 W/mK, which is 40% more accurate than the traditional small - sample training model [57]. In the field of aerospace materials, the machine learning model trained based on multi - features such as molecular weight, chain structure, and cross - linking density has successfully guided the development of new polymer systems with both excellent mechanical strength and thermal stability [75].

The integration of Generative Adversarial Networks (GANs) with coarse-grained molecular dynamics (CGMD) has enabled breakthroughs in material design. For instance, researchers have utilized GANs to generate copolymer structures with targeted Young's modulus, followed by efficient screening via CGMD simulations [6]. Beyond generative models, the TransPolymer model, developed by the Farimani team and based on the Transformer architecture, demonstrates excellent performance in property prediction by effectively capturing polymer sequence and topological features [26]. These data-driven approaches are proving highly effective in application-oriented research. In the field of dielectric materials, for example, machine learning models have accurately predicted the frequency-dependent dielectric behavior of 11,000 unknown polymers, successfully identifying five candidate materials for capacitors and microelectronics applications [71].

The collaborative optimization of Genetic Algorithm and machine learning has greatly improved the efficiency of material development. A study designed 132 new polymers through 100 generations of evolutionary iterations, six of which showed ideal characteristics [76]. The prediction model constructed by Barnett et al. not only accurately evaluated the gas permeability of 700 polymers but also discovered more than 100 excellent materials that exceed the Robeson upper limit [35]. In the design of polyimides, Afzal's team used 29 building blocks to efficiently screen 10,000 candidates from 660 million compounds, and finally obtained the target material with ultra - high refractive index [77].

Autonomous optimization systems have promoted the design of polymer blending systems to enter a new stage. The integrated robot platform, through high - throughput experiments combined with evolutionary algorithms, discovered random heteropolymer blends with performance exceeding that of single components and revealed the regulatory mechanism of molecular fragment interaction on protein thermal stability [78]. Research in the field of thermosetting resins shows that the material genome method combined with machine learning can efficiently design silicon - containing acetylene resins, and high - throughput screening can obtain PSA resins with both excellent processing performance and heat resistance [66]. South Korean scholars innovatively introduced product grade features to construct an XGBoost model, which significantly improved the prediction accuracy of key performance indicators of polymer composites and provided a reliable tool for industrial applications [13].

## 6.2 Optimization of Biodegradable Materials

Machine learning provides a new technical path for the research and development of biodegradable materials, and shows unique advantages especially in performance prediction and structural design. The deep neural network model developed by Bakar's team realizes the accurate prediction of the density characteristics of degradable plastics through principal component analysis and a "coarse - to - fine" optimization strategy [58]. This type of model has excellent non - linear fitting ability and can effectively capture the complex relationship between material structure and performance, showing good stability in the prediction of mechanical properties and degradation behavior. However, it is worth noting that neural networks are more sensitive to data scale. The prediction accuracy will decrease significantly under small sample conditions, and the model interpretability has inherent limitations [58]. In contrast, Support Vector Machines have more advantages in small sample scenarios. The Fransen research group successfully synthesized more than 600 kinds of polyester materials through high - throughput experimental technology combined with machine learning methods, and systematically evaluated their degradation characteristics on Pseudonomas lemoignei [9]. This method constructs a model based on statistical learning theory, which reduces the dependence on large - scale data sets but faces the challenge of computational efficiency when processing massive data.

The inverse design of biodegradable materials is benefiting from the breakthrough of machine learning technology. Researchers have adopted a large - scale screening strategy to generate candidate materials in the design space, and then used the trained prediction model to evaluate their degradation characteristics and mechanical properties [27]. This method has achieved significant results in the optimization combination of natural fibers and bio - based resins. For example, the composite materials of flax, hemp fibers and polylactic acid or polyhydroxyalkanoate recommended by the model have been verified by experiments to show excellent mechanical strength and controllable degradation characteristics under various environmental conditions [75]. Mathematical optimization methods transform material design into a problem of solving objective functions under constraints, and find the optimal solution through deterministic or random algorithms, which effectively alleviates the restriction of combinatorial complexity on design efficiency [27]. The application of transfer learning technology provides a solution to the problem of data scarcity. Studies have shown that pre - trained models have good adaptability in new material systems. For example, Mossa's team successfully applied the convolutional neural network trained on the surfactant system to the perfluorosulfonic acid resin system [23].

At present, the research and development of biodegradable materials still faces key challenges such as accurate regulation of degradation time and optimization of biocompatibility.

Through analyzing the correlation between chemical structure and degradation behavior, machine learning can predict the degradation kinetics characteristics under different environmental conditions [42]. In the research of polylactic acid (PLA)/BiFeO3 (BFO) nanoparticle composite system, BFO was evenly coated on the 3D printed PLA substrate by a simple dip - coating method. The composite system showed excellent piezoelectric photocatalytic degradation performance for Congo Red (CR) and Methylene Blue (MB) (the degradation rates reached 98.9% and 74.3% respectively within 90 minutes). Moreover, with the help of regression models constructed by machine learning models such as Catboost and XGBoost (the $R^2$ values of photocatalysis, piezoelectric catalysis and piezoelectric photocatalysis predictions are 0.93, 0.99 and 0.99 respectively), the application optimization of BFO catalyst was effectively guided, providing a powerful solution for wastewater purification [79]. However, it should be pointed out that the long test cycle of biodegradable materials and the non - uniform experimental standards restrict the construction of high - quality data sets [64]. Future research should focus on the development of multi - scale characterization methods and standardized test schemes to lay a more solid data foundation for the application of machine learning. By integrating automated experimental platforms and high - throughput computing technologies, it is expected to establish a more complete database of biodegradable materials and promote the in - depth development of data - driven design methods in this field [80][81].

### 6.3 Machine Learning Aiding Polymer Material Manufacturing

The field of polymer material manufacturing is experiencing profound changes brought about by machine learning technology, which is rapidly penetrating from laboratory research to industrial practice. The team of Nara Institute of Science and Technology in Japan has made a breakthrough in the research of styrene - methyl methacrylate copolymer system. The flow synthesis method they developed combined with machine learning modeling has significantly improved the mixing effect and heating efficiency [82]. This innovative method not only reduces the time and cost of traditional experiments but also, more importantly, establishes an accurate mathematical model, laying a technical foundation for the industrial production of complex polymer systems.

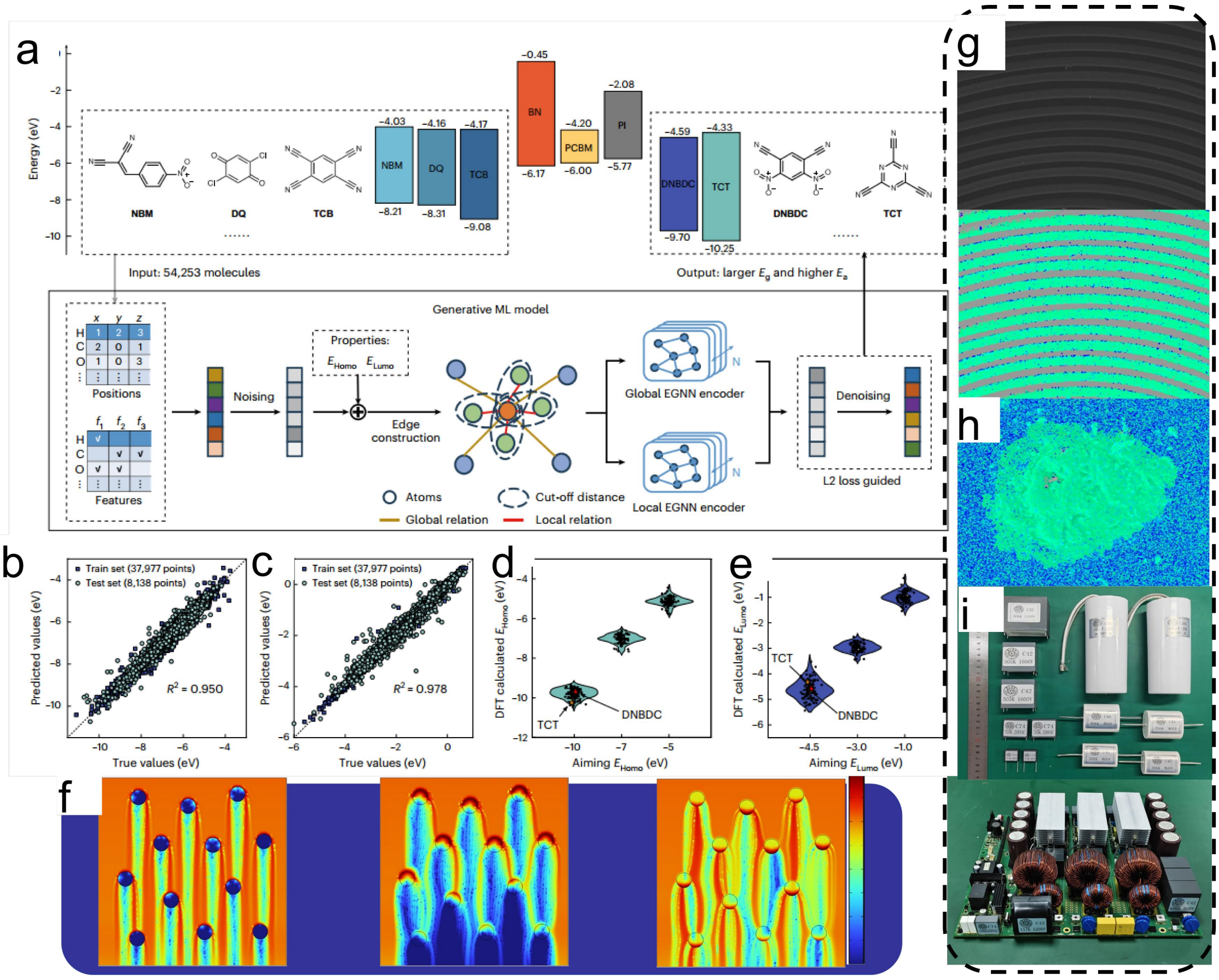

Figure 9 Data-driven development of polymer composite capacitors. (a-f) Model Development & Simulation: (a) Generative model schematic. (b, c) Validation of HOMO/LUMO predictions. (d, e) Benchmarking against DFT calculations. (f) Electron density simulations for different fillers. (g-i) Experimental Characterization: (g) Cross-sectional SEM/EDS of composite. (h) SEM/EDS of a self-healing point. (i) Photographs of final capacitor devices.

Exciting progress has been made in the field of real - time process control. The integration of microfluidic chips and machine learning has achieved a qualitative leap in the regulation of monomer ratio in free radical polymerization (FRP). Studies have shown that this dynamic regulation system can increase the reaction efficiency by 300% [83]. The core of this technology lies in the synergy between online monitoring and machine learning models, which ensures that the reaction process is always in the best state by adjusting process parameters in real time. The injection molding process also benefits from machine learning technology. By in - depth analysis of historical production data, the model can accurately predict key parameters such as mold temperature and cooling rate, thereby effectively avoiding product defects [75]. This predictive method improves product quality while significantly reducing production costs.

The process control of polymer manufacturing is undergoing revolutionary changes brought about by autonomous optimization systems. The newly developed self - driving laboratory platform integrates cloud computing and a variety of online analysis technologies to realize the multi - objective optimization of polymer nanoparticle synthesis [69]. The system can simultaneously optimize multiple performance indicators such as molecular weight distribution and particle size, and find the optimal process parameters through continuous iteration. The NSGA - II algorithm applied in the epoxy resin polymerization process is a successful case, which has achieved significant results in the optimization of number - average molecular weight and polydispersity index [3]. This kind of multi - objective optimization method provides a new idea for solving the balance problem of performance indicators commonly found in industrial production.

Although the application of machine learning in polymer manufacturing has broad prospects, it is still necessary to overcome challenges such as data quality and model generalization. The innovative application of the QLoRA framework provides a new idea for solving the problem of data scarcity. This technology can achieve 91.1% accuracy in processing parameter extraction with only 224 samples [84]. This small - sample learning technology is particularly suitable for scenarios where data acquisition is difficult in polymer manufacturing. Future research should focus on the development of more universal machine learning models and promote the in - depth integration of manufacturing equipment and intelligent algorithms to accelerate the transformation of polymer manufacturing towards comprehensive intelligence.

## 7 Challenges and Prospects

Although the application of machine learning in the field of polymer science has achieved remarkable results, this field still faces a series of technical problems to be solved. As shown in Table 5, the uneven quality of data, insufficient generalization ability of models, and high demand for computing resources have become the main bottlenecks restricting the progress of research. This table systematically summarizes the main technical challenges and their solutions in the machine learning research of polymer materials, including the above - mentioned key issues, and lists the representative solutions and typical cases in the current field, providing methodological references for subsequent research. Especially when dealing with polymer systems with complex structures, the prediction accuracy and stability of existing models are often difficult to meet practical needs. To address these challenges, researchers need to seek breakthroughs from multiple dimensions: constructing a more universal algorithm system, improving the collection and characterization technology of experimental data, and strengthening collaborative innovation between different disciplines. It can be predicted that with the rapid development of high - performance computing technology and the continuous optimization of new algorithms, machine learning will play a more critical role in the field of polymer science, which can not only promote the innovative breakthrough of basic theories but also significantly accelerate the industrialization process of related technologies.

Table 6 Key Technical Challenges and Solutions in Machine Learning Research of Polymer Materials

| Technical Challenge Category | Specific Problem Manifestations | Existing Solutions | Typical Cases/Methods | Citation Source |
|---|---|---|---|---|
| Data Quality | Difficulty in obtaining data of polymer systems, | Constructing standardized databases and integrating multi - source data | FAIR Data sharing program integrates high - throughput experiments, molecular | [7][84] |

| Technical Challenge Category | Specific Problem Manifestations | Existing Solutions | Typical Cases/Methods | Citation Source |
| --- | --- | --- | --- | --- |
| | limited by sample preparation quality | | simulations and literature mining data | |
| Model Generalization Ability | Difficulty in capturing the cross - scale characteristics of polymers (such as chain entanglement, phase separation) | Multi - scale modeling combining physical theory and machine learning | Combining polymer physical theory with machine learning architecture to improve prediction transfer ability | [26][73][89] |
| Computing Resource Requirements | Analyzing high - dimensional data sets consumes a lot of computing resources | GPU - accelerated computing architecture | polyBERT chemical language model uses GPU to improve computing efficiency | [42][44] |
| Model Interpretability | Black - box models are difficult to reveal the intrinsic behavior mechanism of materials | Developing interpretable machine learning methods | Attention mechanism analysis of functional group weight in organic photovoltaic research | [53][87][90] |
| Experimental Verification | Difficulty in digital characterization of complex polymer structures | Automated laboratories and closed - loop optimization systems | NVIDIA ALCHEMI platform realizes the exploration of material chemical space | [69][80][85] |
| Multi - scale Modeling | Single - scale models are difficult to handle multi - scale phenomena of polymers | Developing hybrid multi - scale frameworks | Seamless connection between molecular simulation and continuous - scale machine learning models | [16][89] |

**7.1 Technical Challenges**

Although the introduction of machine learning methods in the field of polymer science has broad prospects, there are still several technical problems to be solved in practical applications. The most urgent problem at present is the difficulty in obtaining high - quality data. High costs and many practical restrictions have seriously restricted the training effect and performance of machine learning models [73]. Taking complex polymer systems as an example, insufficient data makes it difficult for models to accurately capture the cross - scale characteristics of materials, including key features such as random sequences of polymer chains and diversity of condensed state structures [26]. This problem is particularly prominent in the research of solid electrolytes. High - precision molecular simulation methods are difficult to carry out large - scale calculations, while conventional experimental characterization is limited by the quality of sample preparation and cannot effectively distinguish intrinsic ionic conductivity from other interfering factors[84].

The digital characterization of polymer structures also faces severe challenges. Most of the existing characterization methods are limited to the structure of repeating units and cannot fully reflect statistical characteristics such as molecular weight distribution, sequence structure, and topological structure [16]. This characterization defect makes it difficult for machine learning models to fully grasp the complex characteristics of polymer materials. For example, in the study of multi - component polyurethane elastomers, the prediction results of the model on hydrogen bonds on molecular chains are significantly discrete from the overall hydrogen bond distribution of the system, which fully reflects the amorphous characteristics of a single molecular chain in the polymer system [85]. Another tricky problem is the lack of standardized formats for polymer characterization data. Existing data often mixes multiple variables such as molecular weight, processing history, and characterization protocols, which brings great difficulties to data mining and machine learning applications [86].

The lack of model interpretability also limits the in - depth development of machine learning in the polymer field. Traditional AI models generally have the problem of "black box". Although they can produce prediction results, it is difficult to clarify their internal mechanisms [87]. This defect is particularly prominent in fields that need to understand the intrinsic behavior of materials. Taking the research of organic photovoltaic devices as an example, although machine learning methods can accurately model material properties, they often cannot explain which chemical properties play a key role in performance improvement [53]. Another common challenge is the phenomenon of model overfitting, especially when there are many parameters, the model may perform well on the training set, but its prediction ability on new data decreases significantly [88].

Computing resource requirements and algorithm complexity constitute another obstacle. Training large neural networks or analyzing high - dimensional data sets from molecular simulations and spectroscopy consumes a lot of computing resources [42]. When solving the optimal polymer design problem with multi - parameter uncertainty, traditional integration methods will bring a heavy computing burden, and new algorithms need to be developed to deal with this high - dimensional and parameter correlation problem [72]. In addition, existing models perform poorly in dealing with multi - scale phenomena, and the behavioral characteristics of polymer materials often span multiple orders of magnitude, from chain entanglement, phase separation to fracture and creep, but most machine learning tools can only play a role at a single length or time scale [90].

### 7.2 Development Trends

The field of polymer science is experiencing profound changes brought about by machine learning technology, and this change presents the significant characteristics of multi - dimensional and interdisciplinary integration. The data - driven research paradigm is reshaping the pattern of polymer material R & D, among which the combination of multi - scale modeling and physics - informed machine learning methods is particularly striking. The latest research shows that the organic integration of polymer physical theory and machine learning architecture can effectively improve the prediction transfer ability of the model under different conditions, providing a new idea for solving the long - standing problem of complex characterization of polymer systems [73]. The innovation of computing architecture is also worthy of attention. With the iterative upgrading of GPU technology, the computing efficiency of chemical language models such as polyBERT has been significantly improved, making polymer structure design based on molecular fingerprints possible. This full - process automation from prediction to design will completely subvert the traditional trial - and - error research model [44].

In terms of data infrastructure construction, the improvement of standardization and sharing mechanisms has become a consensus in the academic community. At present, polymer data generally faces the problems of chaotic format and uneven quality, and there is an urgent need to establish a unified and standardized data production and analysis process [21]. The advancement of global data sharing programs such as FAIR Data is building a more complete polymer database by integrating multi - source data such as high - throughput experiments, molecular simulations, and literature mining [7]. The continuous expansion of high - quality data sets has significantly improved the prediction accuracy of machine learning models for material performance parameters, especially in key indicators such as the power conversion efficiency of organic photovoltaic devices [53]. The construction of this data ecosystem cannot be separated from the full cooperation of industry, university, and research sectors, and it is necessary to jointly formulate practical data standards and sharing agreements.

The integration of interdisciplinary methods has given birth to the emerging research paradigm of automated laboratories. Cutting - edge research is committed to developing hybrid multi - scale frameworks that combine physical and chemical principles with machine learning algorithms to achieve seamless connection between molecular simulations and continuous - scale machine learning models [89]. The successful development of AI platforms such as NVIDIA ALCHEMI marks that the application of generative AI models in material chemical space exploration and candidate material recommendation has entered the practical stage [80]. The proposal of the concept of autonomous laboratories is more revolutionary. It organically integrates machine learning, robot technology, and cloud computing to build a closed - loop optimization system from material design to synthesis. This integrated innovation has greatly improved R & D efficiency [69].

The development of interpretable machine learning provides a new opportunity for theoretical breakthroughs in polymer science. To address the problem that current black - box models are difficult to reveal internal mechanisms, the academic community is committed to developing more interpretable machine learning methods to make the model decision - making process more transparent [90]. By introducing the knowledge of domain experts and constructing descriptors that can identify the key features of materials, it helps to deeply understand the essential connection between polymer structure and performance [91]. The attention mechanism analysis in the research of organic photovoltaic materials is a typical case. The study found that the model assigns higher weights to adjacent language fragments (usually belonging to the same functional group). This interpretable analysis provides a new perspective for revealing the structure - activity relationship of materials [71]. With the continuous improvement of interpretive tools, machine learning can not only predict material properties but also become an important tool for discovering new scientific laws.

The innovation of the education system has a fundamental supporting role in the development of polymer science. It has become an inevitable choice to integrate programming skills and machine learning knowledge into the chemistry curriculum system. This change aims to cultivate a new generation of polymer scientists with interdisciplinary capabilities [101]. The establishment of industry - university - research collaborative education mechanisms is also crucial. Cultivating compound talents through practical projects can effectively promote the practical application of machine learning technology in the polymer field. This transformation of talent training mode will fundamentally solve the practical dilemma that the threshold of computer majors is too high and synthetic chemists are difficult to apply machine learning tools [92]. It can be predicted that with the continuous deepening of these trends, the application of machine learning in polymer science will achieve a qualitative leap from auxiliary tools to leading paradigms, opening up unprecedented development paths for material innovation.

### 7.3 Application Prospects

The field of polymer science is ushering in profound changes brought about by machine learning technology, and its application potential has penetrated into multiple dimensions such as material R & D, production and manufacturing, and environmental governance. In the development of new materials, the ML - driven workflow is gradually realizing the full - chain automation from literature mining to material synthesis. This closed - loop system compresses the traditional R & D cycle to an unprecedented extent. The design of polymer materials represented by solution polymerized styrene - butadiene rubber (SSBR) has shown the feasibility of machine learning replacing the traditional trial - and - error method, and its accurate prediction ability is expected to be extended to a wider range of material systems and performance indicators [93][94].

The deep integration of the Material Genome Initiative and machine learning is reshaping the methodology of polymer design. Inspired by the breakthrough results of AlphaFold2 in protein structure prediction, deep learning technology provides a new idea for solving the problem of polymer structure prediction. This model has important enlightenment significance for industries such as biopharmaceuticals [95]. Experimental studies have shown that data - driven methods can accurately regulate the morphological characteristics of single - chain nanoparticles (SCNPs), especially under the condition of low functionalization, providing a reliable verification platform for sequence - based design strategies [102]. In the field of mixed matrix membranes, machine learning - assisted high - throughput screening technology significantly improves the performance prediction efficiency of $CO_2$ separation membranes by analyzing the synergistic effect between metal - organic frameworks (MOFs) and polymers [103].

The intelligent transformation of the intelligent manufacturing system cannot be separated from the support of machine learning technology. The development efficiency of materials dedicated to additive manufacturing has achieved a qualitative leap due to data - driven methods.

This innovative model shows unique advantages in addressing material challenges in the fields of bioengineering and aerospace [104]. The newly developed autonomous laboratory platform has built an intelligent optimization system for polymer nano - synthesis by integrating cloud computing and online characterization technology, realizing the accurate production of "on - demand granulation" [105]. It is worth noting that the intelligent information extraction technology based on large language models has made a breakthrough in the optimization of injection molding processes. It can realize the high - precision extraction of processing parameters with only 224 samples, opening up a new way for the digital transformation of traditional manufacturing processes [106].

The development of environment - friendly materials is achieving leapfrog development with the help of machine learning. The performance optimization research of materials such as polyurethane elastomers reveals that the eigenvalue of system parameters has a more significant impact on material performance than the characteristics of a single molecular chain. This finding provides an important basis for the overall regulation of material components [107]. The discovery of ionene materials with Troger's base structure marks a major progress in the field of sustainable energy materials. Their excellent conductivity provides an innovative idea for the design of a new generation of lithium - ion batteries [108]. Biomimetic intelligent thermal management materials, by simulating the biological thermal regulation mechanism and combining machine learning optimization strategies, show the unique value of dynamic regulation in fields such as wearable devices and building energy conservation [109].

The integration of interdisciplinary technologies continues to expand the application depth of machine learning in polymer science. The polyBERT chemical language model provides a universal solution for polymer space exploration with its high - throughput screening capability [110]. Graph Neural Networks (GNNs) perform well in capturing the topological structure information of polymer chains, establishing a new paradigm for molecular ensemble modeling [111]. With the iterative upgrading of professional computing platforms, AI proxy models such as Machine Learning Interatomic Potentials (MLIPs) will accelerate the process of material discovery, promote the paradigm shift of polymer science from experience - driven to data - driven, and finally realize the historic leap of material R & D from "trial - and - error game" to "precision navigation" [112].